%% file: final_manuscript.tex
\documentclass[12pt, draftclsnofoot, onecolumn]{IEEEtran}

\usepackage{cite}
\usepackage{graphicx,amsmath,amssymb}
\usepackage{subfigure}
\usepackage{citesort}
\usepackage{fancyhdr}
\usepackage{mdwmath}
\usepackage{mdwtab}
\usepackage{balance}
\usepackage{xcolor}
\usepackage{bm}

\usepackage{algorithm}
\usepackage{multirow}
\usepackage{flafter}
\usepackage{comment}

\usepackage{amsthm}
\usepackage{algpseudocode}

\usepackage{stmaryrd}

\usepackage[font=small]{caption}
\usepackage{mathtools}
\newtheorem{remark}{Remark}
\newtheorem{theorem}{Theorem}
\newtheorem{lemma}{Lemma}
\newtheorem{model}{Model}

\newtheorem{corollary}{\textbf{Corollary}}

\allowdisplaybreaks
\usepackage{float}
\floatstyle{plaintop}
\restylefloat{table}

\newcommand\myeq{\stackrel{\mathclap{\normalfont\mbox{(a)}}}{=}}
\begin{document}

%\title{Fairness of User Clustering in MIMO Non-orthogonal Multiple Access Systems}
%\author{Author 1,  Author 2, Author 3, and Author 4}
%%\markboth{IEEE Transactions on \LaTeX\ }
%%{Hayes}

\title{Exploiting STAR-RISs in Near-Field Communications}
%\title{Modeling and Analysis of Metasurface-Based STAR-RISs}
%

\author{Jiaqi\ Xu,~\IEEEmembership{Graduate Student Member,~IEEE}, Xidong\ Mu,~\IEEEmembership{Member,~IEEE}, and Yuanwei\ Liu,~\IEEEmembership{Senior Member,~IEEE}

\thanks{J. Xu, X. Mu, and Y. Liu are with the School of Electronic Engineering and Computer Science, Queen Mary University of London, London E1 4NS, UK. (email:\{jiaqi.xu, xidong.mu, yuanwei.liu\}@qmul.ac.uk).}

}
\maketitle
%\thispagestyle{fancyplain}
%\pagestyle{fancy}
\begin{abstract}
The reconfigurable intelligent surface (RIS) is a promising technology to provide smart radio environment.
In contrast to the well-studied patch-array-based RISs, this work focuses on the metasurface-based RISs and simultaneously transmitting and reflecting (STAR)-RISs where the elements have millimeter or even molecular sizes.
For these meticulous metasurface structures, near-field effects are dominant and a continuous electric current distribution should be adopted for capturing their electromagnetic response instead of discrete phase-shift matrices.
Exploiting the electric current distribution, a Green's function method based channel model is proposed.
Based on the proposed model, performance analysis is carried out for both transmitting/reflecting-only RISs and STAR-RISs.
1) For the transmitting/reflecting-only RIS-aided single-user scenario, 
closed-formed expressions for the near-field/far-field boundary and the end-to-end channel gain are derived. 
%Then, the number of communication modes, i.e.,
Then, degrees-of-freedom (DoFs) and the power scaling laws are obtained. It is proved that the near-field channel exhibits higher DoFs than the far-field channel. 
It is also confirmed that when communication distance increases beyond the field boundary, the near-field power scaling law degrades to the well-known far-field result.
2) For the STAR-RIS-aided multi-user scenario, three practical STAR-RIS configuration strategies are proposed, namely power splitting (PS), selective element grouping (SEG), and random element grouping (REG) strategies.
The channel gains for users are derived within both the pure near-field regime and the hybrid near-field and far-field regime.
Finally, numerical results confirm that: 1) for metasurface-based RISs, the field boundary depends on the sizes of both the RIS and the receiver, 2) the received power scales quadratically with the number of elements within the far-field regime and scales linearly within the near-field regime, and 3) for STAR-RISs, SEG has the highest near-field channel gain among the three proposed strategies and PS yields the highest DoFs for the near-field channel.
%for the multi-user case in mixed near-field and far-field regimes, the signal power of the near-field user can be improved by about $5$ dB compared to the case without STAR-RIS and $3$ dB compared to the case without passive STAR-RIS beamforming.
\end{abstract}

\section{Introduction}
The sixth-generation (6G) wireless communication networks are envisioned to deliver significant performance improvements beyond existing technologies~\cite{9598915}. To sustainably achieve these goals, a promising way is to rely on new physical layer technologies. Recently, reconfigurable intelligent surfaces (RISs) have been proposed~\cite{di2019smart}. Consisting of a large number of passive reflective elements, an RIS can dynamically adjust the phase-shift of the reflected signals. Specifically, owning to the remarkable development of metasurface technologies over the past decade, it is now possible to accurately control the electromagnetic signal on sub-wavelength scale~\cite{chen2016review}. Thus, wavefront shaping for radio-frequency waves, millimeter waves, and even optical waves has become possible exploiting RISs with different period sizes~\cite{ahead}.
To further extend the coverage of RISs to full-space, a novel category of simultaneously transmitting and reflecting (STAR)-RISs has been proposed~\cite{xu_star}. By facilitating both transmission and reflection of the incident signals on each element, STAR-RISs can simultaneously reconfigure the channels of multiple users located on different sides of the surface. This also removes the unnecessary topological constraint which often appears in the deployment of transmitting/reflecting-only RISs~\cite{liu_star}.
Due to the large aperture size of RISs and STAR-RISs, their near-field regions are significantly larger than conventional transmitters, especially for cases where millimeter-wave (mmWave) and terahertz (THz) frequency bands are used.
The smart radio environment coverage enabled by conventional RISs and STAR-RISs has attracted significant research interests. 
Among these research activities, the channel modeling and
performance analysis of conventional RISs and STAR-RISs are the most fundamental aspects.
In the following, we briefly introduce these existing works.

\subsection{Prior Works}
\subsubsection{Conventional RISs}

Depending on the location of the receiver, wireless communication might occur in the far-field regime or the near-field regime of an RIS. The boundary between these two regimes is usually considered to be $2L^2/\lambda$, where $L$ is the largest dimension of the RIS and $\lambda$ is the wavelength of the carrier signal. 
The far-field performance of reflecting-only RISs has been thoroughly studied in the pioneering studies~\cite{ahead,zhangrui}. Recently, increasing research interests have been focusing on the near-field performance of RISs or intelligent reflecting surfaces (IRSs). In~\cite{nearfield}, the near-field power scaling law and finite asymptotic SNR limits were derived. In~\cite{ajam2022power}, the power scaling law for optical IRSs was studied. In~\cite{9139337}, the achievable near-field degree-of-freedom (DoF) and gain for large intelligent surfaces (LISs) were derived. It is proved that the Friis' formula is no longer valid within the near-field regime.

\subsubsection{STAR-RISs}
Motivated by the aforementioned favorable characteristics of STAR-RISs, extensive research efforts have been devoted to exploiting STAR-RIS in existing wireless technologies, such as non-orthogonal multiple access (NOMA)~\cite{zuo2021joint}, full-duplex communications~\cite{perera2022sum}, and federated learning~\cite{9815289,9685556}.
For STAR-RIS-aided wireless communication within the far-field regime, existing works modeled patch-array-based STAR-RISs using transmission and reflection (T\&R) coefficients. 
In~\cite{xu_star} and \cite{xu_correlated}, independent and coupled phase-shift models was proposed for STAR-RISs.
For the optimization and performance analysis, existing works optimized the far-field multi-user communication for different objectives. 
In \cite{mu2021simultaneously}, the active beamforming at the base station and the passive beamforming at the STAR-RIS were studied for minimizing the transmit power with three proposed STAR-RIS operating protocols. In \cite{star_coverage}, the fundamental coverage performance of STAR-RIS-assisted communications was investigated. In \cite{zuo2021joint}, the STAR-RIS-assisted multi-user NOMA communication system was studied where a suboptimal two-layer iterative algorithm was proposed for maximizing the sum rate.
For the STAR-RIS-aided wireless communication in the near-field regime, \cite{xu_star} proposed a near-field channel model based on the Huygens–Fresnel principle and the Kirchhoff's diffraction formula.

\subsection{Motivations and Contributions}

\begin{table}[!t]
\small
\centering
\begin{tabular}{|c|c|cc|}
\hline
\multirow{2}{*}{Implementations}   & \multirow{2}{*}{Field Region} & \multicolumn{2}{c|}{Research Contributions}                                                             \\ \cline{3-4} 
                                   &                               & \multicolumn{1}{c|}{RISs}                                & STAR-RISs                                    \\ \hline
\multirow{2}{*}{Patch-Array-Based} & Far-Field                     & \multicolumn{1}{c|}{\cite{di2019smart,zhangrui}} & \cite{liu_star,xu_correlated,mu2021simultaneously} \\ \cline{2-4} 
                                   & Near-Field                    & \multicolumn{1}{c|}{\cite{ahead,nearfield,ajam2022power}}   & \cite{xu_star}             \\ \hline
\multirow{2}{*}{Metasurface-Based} & Far-Field                     & \multicolumn{2}{c|}{\multirow{2}{*}{This work}}                                                         \\ \cline{2-2}
                                   & Near-Field                    & \multicolumn{2}{c|}{}                                                                                   \\ \hline
\end{tabular}
\caption{Comparing the scope of this work with existing representative works.}
\label{tab:comp}
\end{table}

Note that hardware implementations of RISs and STAR-RISs can be classified into two major categories, namely patch-array-based RISs and metasurface-based RISs~\cite{xu_vtmag}. Specifically, patch-array-based RISs/STAR-RISs consist of periodic cells having sizes on the order of a few centimetres and the cells can be made tunable by accommodating diodes or delay lines. By contrast, matasurface-based RISs/STAR-RISs have cells on the order of a few millimetres, or even molecular sizes.
While various research contribution focused on patch-array-based RISs and STAR-RISs~\cite{liu_star,xu_correlated,mu2021simultaneously,wu_estimation,star_coverage,zuo2021joint,perera2022sum,9815289,9685556}, the modeling and analysis of metasurface-based RISs and STAR-RISs are still in their infancy. 
Due to their small period structures, the metasurface-based RISs and STAR-RISs can be employed for wireless communication in high carrier frequencies~\cite{ahead,xu_vtmag}, such as millimeter wave (mmWave) and terahertz (THz) frequency bands. As a result, their near-field effects are more dominated than their patch-array-based counterparts (the near-field regime is larger).
However, the aforementioned works on near-field analysis~\cite{xu_star,ajam2022power,nearfield} mainly focused on the patch-array-based RISs. As summarized in Table~\ref{tab:comp}, there is a lack of comprehensive study on the channel modeling of metasurface-based RISs and STAR-RISs, which provides the main motivations of this work.
%More importantly, the channel models used for RISs~\cite{nearfield} and STAR-RISs~\cite{xu_correlated} are based on the phase-shift matrix. These modeling methods are not applicable to near-field STAR-RISs because they does not take into account the correlation between the transmitted and reflected signals. Besides, modeling STAR-RISs through its transmission and reflection phase-shift coefficient matrices is limited to the patch-array implementation, and thus, cannot provide the fundamental performance limit for STAR-RISs implemented by smart glasses or metasurfaces~\cite{xu_vtmag}.

To reveal the fundamental performance limit of metasurface-based RISs and STAR-RISs within both the far-field and the near-field regimes, we propose a general modelling method by characterizing the RIS elements using a continuous current distribution within each element. By exploiting this current-based characterization of RISs and STAR-RISs, we propose a channel model based on the Green's function method. The channel modeling is carried out by integrating the contribution for all induced currents within the volume of the receiver. In contrast to the phase-shift characterization, this channel modeling method reveals the best-achievable RIS-aided channel gain between the transmitter and receiver in both the near-field and far-field regimes. The main contributions of this work can be summarized as follows:
\begin{itemize}
    \item We propose a novel channel model for metasurface-based RISs and STAR-RISs, where we exploit the Green's function method and the current-based characterization for the elements. We illustrate how transmitting-only RISs, reflecting-only RISs, and STAR-RISs can be achieved by configuring the current distributions. We then analyze the near-field and far-field performance in both the transmitting/reflecting-only RISs-aided single-user scenario and the STAR-RIS-aided multi-user scenarios. 
    \item For transmitting/reflecting-only RIS-aided single-user scenario, we derive the position of the field boundary by taking into account the size of both the RIS and the receiver. We also derive the fundamental performance limits of RIS-aided communication in terms of the maximum end-to-end channel gain, DoF, and the power scaling law. These obtained near-field metrics are then compared with those in the far-field regime. Specifically, we show that the received power scales linearly with the number of elements within the near-field regime and scales
    quadratically within the far-field regime.
    \item For STAR-RIS-aided multi-user scenario, we propose three practical configuration strategies for employing STAR-RISs, namely power splitting (PS), selective element grouping (SEG), and random selective element grouping (REG) strategies. We derive closed-form expressions for the channel gain of each proposed strategies within the pure near-field regime. We also evaluate the performance of STAR-RIS in the hybrid near-field and far-field regime, where the STAR-RIS is deployed to assist one near-field user on the transmission side and one far-field user on the reflection side. The channel gains for both users are derived for the case without STAR-RIS and for the case with STAR-RIS under three proposed configuration strategies.
    % Numerical results reveal that the STAR-RIS improves the received power of the indoor user by about $5$ dB compared to the case without STAR-RIS and $3$ dB compared to the case with an open window (without passive STAR-RIS beamforming).}
    \item Our simulation results verify that 1) for metasurface-based RISs with larger sizes and receivers with larger aperture areas, the near-field regime expands and the field boundary is further away from the RIS, 2) for the STAR-RIS-aided multi-user pure near-field communication scenario, the SEG strategy has the best sum rate performance of all three strategies but yields lower DoFs for the end-to-end channel, and 3) for the STAR-RIS-aided multi-user hybrid near- and far-field communication scenario, STAR-RIS can significantly increase the channel gains of users and thus can improve the signal coverage especially in the indoor ``blind zone''.
\end{itemize}

\subsection{Organization}
The rest of this paper is organized as follows: Section II introduces the Green's function method, based on which the channel model for metasurface-based RISs and STAR-RISs is proposed.
Section III analyze the fundamental performance limits in the transmitting/reflecting-only RIS-aided single-user scenario. The upper bound of the end-to-end channel gain, DoF, and the power scaling law are given. 
In Section IV, the three proposed configuration strategies for STAR-RIS-aided multi-user communication scenario are given. The performance analysis is carried out in both pure near-field regime and hybrid near- and far-field regime. Finally, Section V provides numerical results to verify our theoretical analysis.

\section{A Green's Function Method Based Channel Model for Metasurface-Based RISs and STAR-RISs}\label{sec_hardware}

In this section, we give the general formulation of the Green's function method based channel modeling method for metasurface-based STAR-RISs, where transmitting/reflecting-only RISs can be regarded as a special case. Then, we present the current-based characterization of STAR elements and the channel model for both near-field and far-field regimes. We then show the differences and connections between the proposed model and 
the transmission and reflection (T\&R) coefficients based channel model for patch-array-based
STAR-RISs, which was previously given in~\cite{xu_star}. Finally, we give an example on how reflecting-only RISs, transmitting-only RISs, and STAR-RISs can be modeled with different equivalent current distributions.

\subsection{The Green's Function Method}
We denote the volume (space) occupied by the STAR-RIS (or RIS) as $V_T$ and the volume of the receiver as $V_R$. The EM response of STAR-RIS can be characterized using equivalent current distribution within $V_T$.
According to electromagnetic (EM) theory, the reflected/transmitted EM field in the region outside of $V_T$ can be calculated using the electric current $\mathbf{J}(\mathbf{r}')$, where $\mathbf{r}'$ is the position of the source point within $V_T$. Particularly, exploiting the Green's function approach, the electric field vector at field point $\mathbf{r}$ (position of the receiver) is given by~\cite[Eq. (9)]{danufane2020path}:
\begin{equation}\label{erg}
    \mathbf{E}(\mathbf{r}) = \int_{V_T} \overline{\overline{G}}(\mathbf{r},\mathbf{r}') \mathbf{J}(\mathbf{r}') \mathrm{d}V_T,
\end{equation}
where $\overline{\overline{G}}(\mathbf{r},\mathbf{r}')$ is the tensor Green's function. The exact form of $\overline{\overline{G}}(\mathbf{r},\mathbf{r}')$ depends on the boundary conditions and the position of the user~\cite{danufane2020path}. If multiple users are involved, each of them corresponds with a different Green's function.
%In \eqref{erg}, the Green's function can be regard as an operator which maps $\mathbf{J}(\mathbf{r}')$ to $\mathbf{E}(\mathbf{r})$. We elaborate on the details of this linear mapping and the matrix representation of $\overline{\overline{G}}(\mathbf{r},\mathbf{r}')$ in Appendix~\ref{appendix}.

\subsection{Current-Based Characterization of STAR-RIS Elements}
In this work, the STAR-RIS is characterized using the strength and phases of the equivalent current, i.e., $\mathbf{J}(\mathbf{r}')$. Its distribution depends on the incident field and the surface impedance configuration of the STAR-RIS elements. In prior works, the patch-array-based STAR-RIS is modeled using the transmission and reflection (T\&R) coefficients of each element~\cite{xu_star,xu_correlated}. Note that the T\&R coefficient of each patch is the macroscopic effect of the microscopic electric currents induced within each STAR-RIS element. For the purpose of obtaining fundamental performance limit for metasurface-based STAR-RIS, in this work, we use the equivalent current-based characterization. However, the T\&R coefficients characterization for the $m$th element can be related to the equivalent current-based characterization through the following integration:
\begin{align}\label{tm}
    T_m = \frac{|\mathbf{E}(\mathbf{r}_T)|}{|\mathbf{E}_{inc}|} = \left|\int_{V_m} \overline{\overline{G}}(\mathbf{r}_T,\mathbf{r}') \mathbf{J}(\mathbf{r}') \mathrm{d}V_m\right|,\\ \label{rm}
    R_m = \frac{|\mathbf{E}(\mathbf{r}_R)|}{|\mathbf{E}_{inc}|} =\left| \int_{V_m} \overline{\overline{G}}(\mathbf{r}_R,\mathbf{r}') \mathbf{J}(\mathbf{r}') \mathrm{d}V_m\right|,
\end{align}
where $|\mathbf{E}|$ denotes the complex amplitudes of a field, $\mathbf{r}_T$ and $\mathbf{r}_R$ are the reference point on the transmission/reflection side of the STAR-RIS, $\mathbf{J}(\mathbf{r}')$ is the electric current induced by a normalized incident field, and $V_m$ is the volume of the $m$th element.
In practice, the induced current $\mathbf{J}(\mathbf{r}')$ can be configured by adjusting the electric impedance of the STAR-RIS elements~\cite{zhu2014dynamic}. Thus, using the equivalent current instead of the T\&R coefficients, we can obtain more fundamental performance limits of STAR-RISs\footnote{The current distribution provides a more detailed EM-response than the element-wise T\&R phase-shift coefficients. The phase-shift model uses only a single phase-shift value for the entire element, thus ignoring the phase difference of the electric fields within each element. Also, in existing models, the phase-shift coefficients are assumed to be the same for receivers in different directions, although this is not the case in practice.}.

\subsection{Channel Model}

\begin{figure}[t!]
    \begin{center}
        \includegraphics[scale=0.35]{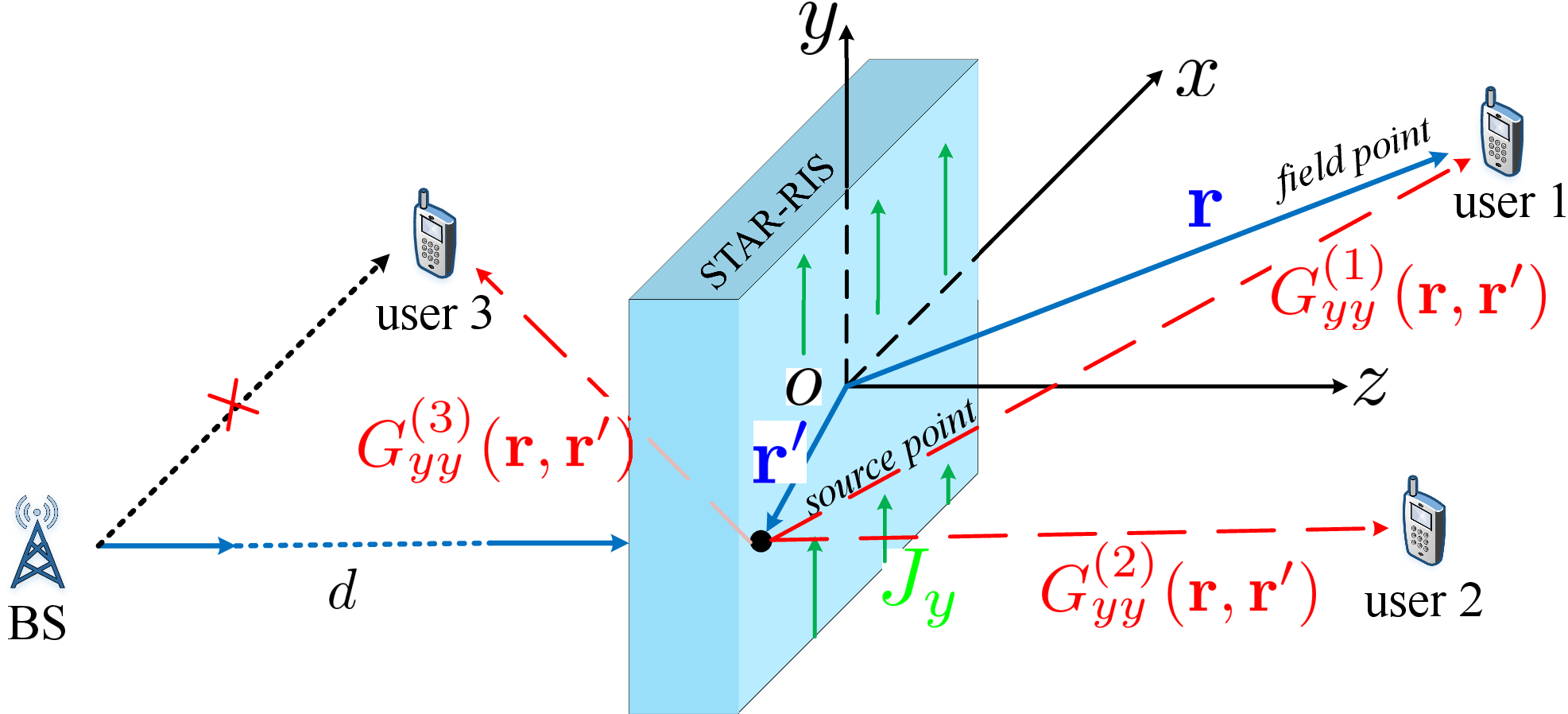}
        \caption{A Green's function-based channel model for metasurface-based STAR-RISs.}
        \label{system_multi}
    \end{center}
\end{figure}

As illustrated in Fig.~\ref{system_multi}, we consider a STAR-RIS centered at the origin of the coordinate. With the assistance of a STAR-RIS, a single antenna base station (BS) communicates with receivers which are located on the $x-z$ plane. We assume that the BS is located in the far-field of the STAR-RIS and the receivers are within the near-field of STAR-RIS. Consider the case of a vertically polarized wireless signal, the equivalent electric currents induced within $V_T$ are in the $y$-direction, i.e., $\mathbf{J}(\mathbf{r}')=\hat{\mathbf{y}}J_y(\mathbf{r}')$, and the electric field they generate at receiver $i$ is $\mathbf{E}^{(i)}(\mathbf{r}) = \hat{\mathbf{y}}E^{(i)}_y(\mathbf{r})$. According to \eqref{erg}, these received fields are given by:
\begin{equation}\label{ey}
    E^{(i)}_y(\mathbf{r}) = \int_{V_T} G^{(i)}_{yy}(\mathbf{r},\mathbf{r}') J_y(\mathbf{r}') \mathrm{d^3}\mathbf{r}',
\end{equation}
where $G^{(i)}_{yy}$ is the $(y,y)$-element of the Green's tenser for receiver $i$. For free space transmission, we have~\cite{rothwell2018electromagnetics}:
\begin{equation}\label{gyy}
    G^{(i)}_{yy}(\mathbf{r},\mathbf{r}') = -(j\omega\mu_0 + \frac{k^2}{j\omega\epsilon_0})\frac{\exp{(-jk|\mathbf{r}-\mathbf{r}'|)}}{4\pi|\mathbf{r}-\mathbf{r}'|},
\end{equation}
where $\mu_0$ and $\epsilon_0$ are the free space permeability and permittivity, respectively, $\omega$ and $k$ are the signal frequency and wave number, respectively.

Next, we formulate the received power at the receiving volume $V_R$. To obtain the largest received power possible, $J_y(\mathbf{r}')$ must be configured such that the electric field it generate, according to \eqref{ey}, yield the largest integrated E-field magnitude within $V_R$, i.e.,
\begin{equation}\label{A}
    |h_{\text{T-to-R}}|^2 = \int_{V_R} E_y^*(\mathbf{r})E_y(\mathbf{r})\mathrm{d^3}\mathbf{r}.
\end{equation}
For brevity, we omit the index $(i)$ for the $i$th user and give the end-to-end channel gain between the BS and $V_R$ in the following lemma:
\begin{lemma}
The maximum magnitude of the end-to-end channel gain between the BS and the receiving volume ($V_R$) is given as follows:
\begin{equation}\label{A_2}
    |h_{\text{end-to-end}}|^2 = \frac{DA_T}{4\pi d^2}\int_{V_T} J^*_y(\mathbf{r}'_1) \int_{V_T} K(\mathbf{r}'_1,\mathbf{r}'_2)J_y(\mathbf{r}'_2)\mathrm{d^3}\mathbf{r}'_1 \mathrm{d^3}\mathbf{r}'_2,
\end{equation}
where $D$ is the directivity of the BS antenna, $A_T$ is the aperture size of STAR-RIS facing the base station, $d$ is the distance between BS and STAR-RIS, $J_y(\mathbf{r})$ is the normalized current distribution within $V_T$, i.e., $<\mathbf{j}_1,\mathbf{j}_2> = \int_{V_t} \mathbf{j}^T_1\mathbf{j}^*_2\ dV_T = 1$, $\mathbf{A}^T$ is the transpose of vector $\mathbf{A}$, $^*$ denotes the complex conjugate, and
\begin{equation}\label{K_def}
    K(\mathbf{r}'_1,\mathbf{r}'_2) = \int_{V_R} G^*_{yy}(\mathbf{r},\mathbf{r}'_1)G_{yy}(\mathbf{r},\mathbf{r}'_2)\mathrm{d^3}\mathbf{r}
\end{equation}
is a kernel function in the space within $V_T$.
\begin{proof}
The channel gain can be expressed as the product of two parts, $D\frac{A_T}{4\pi d^2}$ is the maximum channel gain between BS and the STAR-RIS, and $|h_{\text{T-to-R}}|^2$ is the gain between STAR-RIS and a receiver. For the first part, according to the Friis' formula, the received power from a far-field antenna with a directivity of $D$ is $P_{r} = D\frac{A_T}{4\pi d^2} P_{t}$, where $P_{t}$ is the transmit power.
For the second part, the proof is straightforward by substituting \eqref{ey} into \eqref{A}:
\begin{equation}\label{int3}
    |h_{\text{T-to-R}}|^2 = \int_{V_R}\int_{V_T}\int_{V_T} G^*_{yy}(\mathbf{r},\mathbf{r}'_1) J_y(\mathbf{r}'_1) G_{yy}(\mathbf{r},\mathbf{r}'_2) J_y(\mathbf{r}'_2) \mathrm{d^3}\mathbf{r}'_1 \mathrm{d^3}\mathbf{r}'_2\mathrm{d^3}\mathbf{r}
\end{equation}
Since in \eqref{int3}, only the Green's functions $G_{yy}$ are functions of the field point $\mathbf{r}$, the integral within $V_R$ can be evaluated separately as a kernel, i.e., $ K(\mathbf{r}'_1,\mathbf{r}'_2)$.
\end{proof}
\end{lemma}

\begin{remark}

The kernel function can be understand as a measurement of the coupling strength of two source point (located at $\mathbf{r}'_1$ and $\mathbf{r}'_2$) which is then averaged within $V_R$.
According to its definition, $K(\mathbf{r}'_1,\mathbf{r}'_2)$ is a Hermitian operator, i.e., $K(\mathbf{r}'_1,\mathbf{r}'_2) = K^*(\mathbf{r}'_2,\mathbf{r}'_1)$, and thus is diagonalizable under a complete set of basis functions in $V_T$. According to linear algebra~\cite{porter1990integral}, the quantity in \eqref{A_2} is maximized when $J_y(\mathbf{r}')$ is chosen to be the eigenfunction of $K(\mathbf{r}'_1,\mathbf{r}'_2)$ with the largest eigenvalue. In other words, we have the following relation for $h_{\text{T-to-R}}$, $J_y(\mathbf{r}')$, and $K(\mathbf{r}'_1,\mathbf{r}'_2)$:
\begin{equation}\label{eigen}
    |h_{\text{T-to-R}}|^2\cdot J_y(\mathbf{r}'_1) =  \int_{V_T} K(\mathbf{r}'_1,\mathbf{r}'_2)J_y(\mathbf{r}'_2)\mathrm{d^3}\mathbf{r}'_2.
\end{equation}
In practice, $J_y(\mathbf{r}')$ is determined by the phase-shift configuration of the STAR-RIS elements and the kernel $K(\mathbf{r}'_1,\mathbf{r}'_2)$ is determined by the positional information of the receiver as well as the environment.
In addition, the DoF of the end-to-end channel is related to the number of eigenfunctions of the kernel function in \eqref{eigen} with substantially large eigenvalues~\cite{miller2000communicating}. In the following sections, we can obtain the closed-form channel gain by choosing different current distribution $J_y(\mathbf{r}')$ for transmitting/reflecting-only RISs and STAR-RISs.
\end{remark}

\subsection{Modeling of Reflecting-Only RISs, Transmitting-Only RISs, and STAR-RISs with Equivalent Current Distributions}\label{added_sec}
Before commencing with deriving the closed-form channel gain for the metasurface-based RISs and STAR-RISs, we elaborate on the relation between the distribution of $J_y(\mathbf{r}')$ and the types of RISs, i.e., reflecting-only RISs, transmitting-only RISs, and STAR-RISs.
As shown in \eqref{tm} and \eqref{rm}, the T\&R coefficients are determined by the distribution of $J_y(\mathbf{r}')$ within each macroscopic element ($V_m$). To give the readers a better idea of the Green's function-based channel model, we give examples of the exact distributions of $J_y(\mathbf{r}')$ which corresponds to the different RIS types.

\begin{figure*}[t!]
\centering
\subfigure[Transmitting-Only]{\label{n01t}
\includegraphics[width= 1.9in]{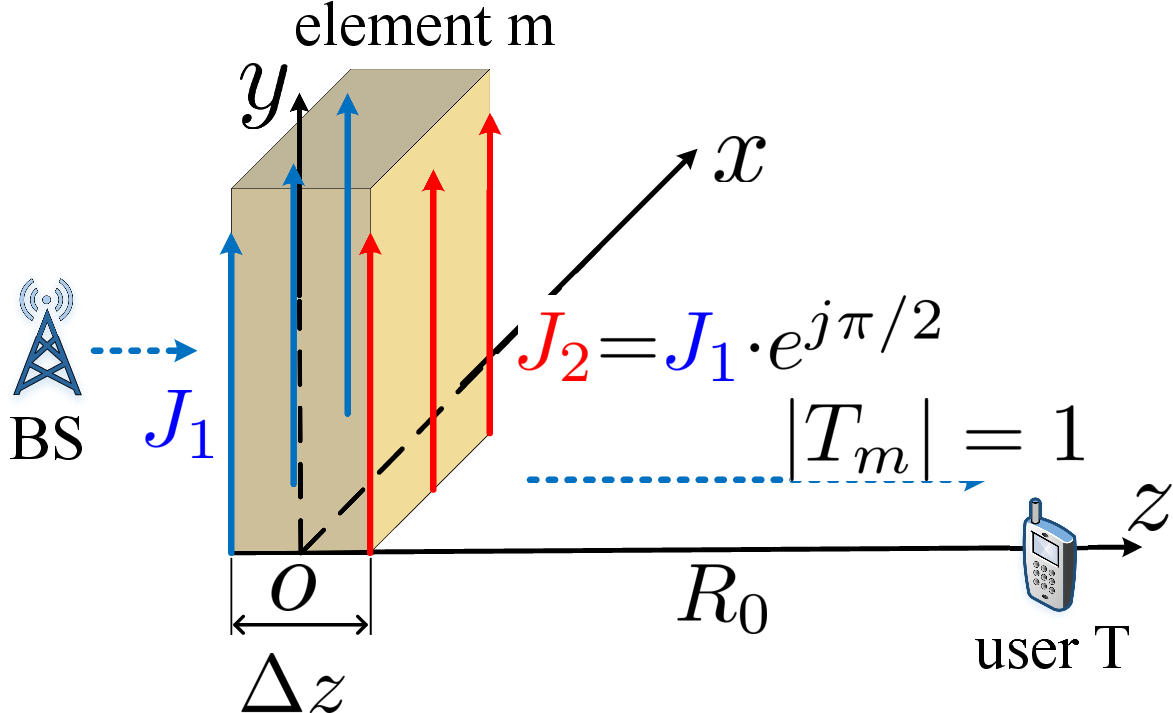}}
\subfigure[Reflecting-Only]{\label{nb1t}
\includegraphics[width= 1.9in]{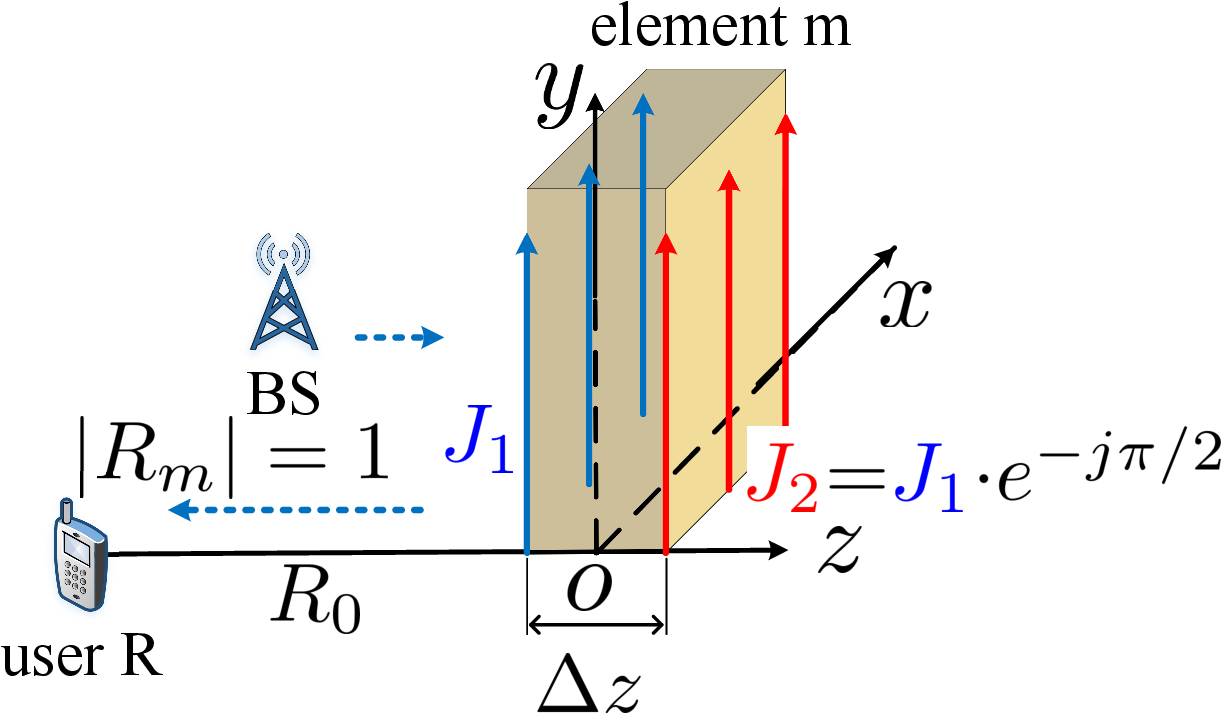}}
\subfigure[STAR-RIS]{\label{na1t}
\includegraphics[width= 1.9in]{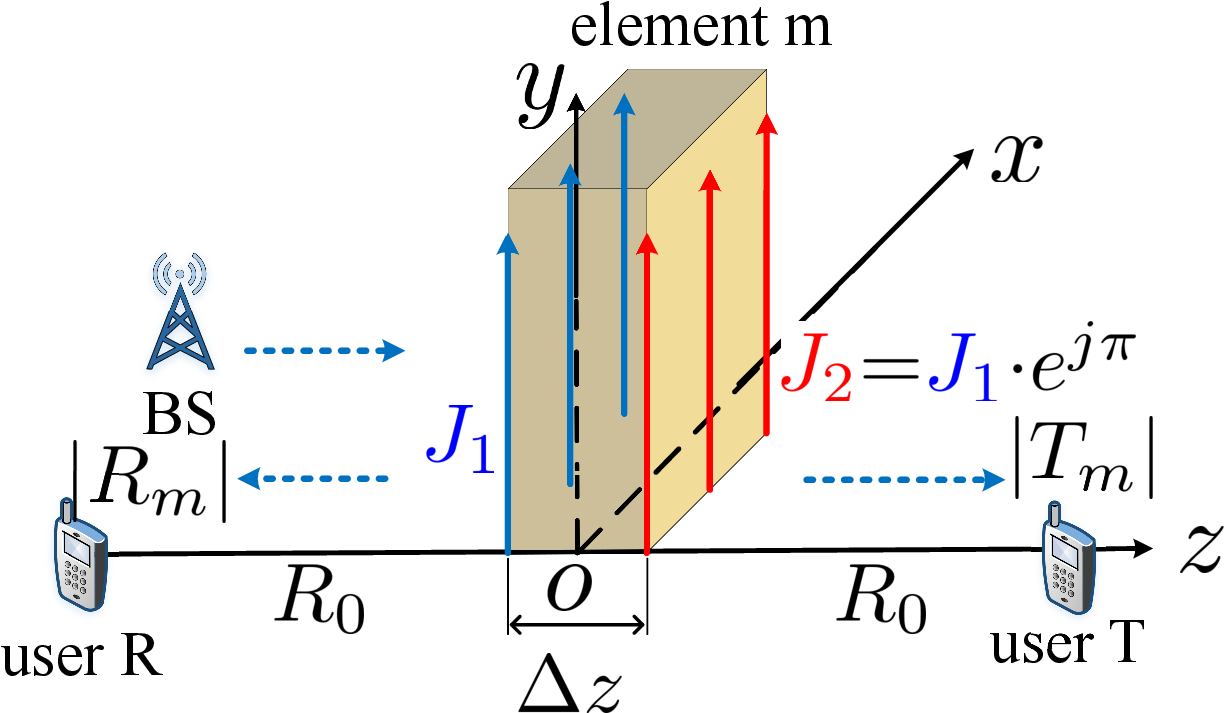}}
\caption{Illustration of current distributions for different RIS types.}\label{nicet}
\end{figure*}

As shown in Fig.~\ref{nicet}, the considered element is centered at the origin and one/two receivers are located on the $z$-axis with distances $R_0$ to the element\footnote{For illustration purposes, the dimensions of this figure is not to scale.}.
Assume that the width of the element $\Delta z = \lambda/4$ and the size of the element is small compared to $R_0$. Thus, according to the kernel function can be reduced to a function of $z'$, i.e., $G_{yy}(z') = \beta^2 e^{-jk|R_0-z'|}/(4\pi |R_0-z'|)$, where $\beta = j\omega\mu_0 + \frac{k^2}{j\omega\epsilon_0}$. In this example, we consider that the induced currents are only present on the front and back surfaces of the element. As illustrated in Fig.~\ref{nicet}, we denote the surface density of the currents on the incident side and on the transmission side as $J_1$ and $J_2$, respectively. Under these assumptions, the integration in \eqref{tm} and \eqref{rm} simply reduce to two-sum:
\begin{align}
    T_m = J_1\cdot G_{yy}\left(|R_0+\Delta z/2|\right) + J_2\cdot G_{yy}\left(|R_0-\Delta z/2|\right),\\
    R_m = J_1\cdot G_{yy}(|R_0-\Delta z/2|) + J_2\cdot G_{yy}(|R_0+\Delta z/2|).
\end{align}
Note that both $J_1$, $J_2$, and $G_{yy}$ are complex values with amplitudes and phases. Exploiting the fact that $R_0 \gg \Delta z$ and $e^{jk\Delta z} = e^{j\pi/2}$, the macroscopic T\&R coefficients of this element can be further simplified as:
\begin{equation}\label{tmrm}
    T_m = G_0\left(J_1\cdot e^{j\pi/2} + J_2 \right) \text{, \ and,} \ R_m = G_0\left(J_1 + J_2\cdot e^{j\pi/2} \right),
\end{equation}
where $G_0 = \beta^2 e^{|R_0-\Delta z/2|}/(4\pi R_0)$.
Here, \eqref{tmrm} reveals the relation between the surface current configuration, i.e., $J_1$, $J_2$ and the T\&R coefficients. Suppose that the surface currents have the same amplitudes and we rewrite them as $J_1 = |J_0| e^{j\phi_1}$ and $J_2 = |J_0| e^{j\phi_2}$. The reflecting-only RISs, transmitting-only RISs, and STAR-RISs can be modeled by the phase differences of the currents, i.e., $\phi_1-\phi_2$, according to Table~\ref{tab:my-table1}.
\begin{table}[!h]
\centering
\small
\begin{tabular}{|c|c|c|c|}
\hline
RIS Types         & Transmitting-Only                                & Reflecting-Only & STAR \\ \hline
Configuration of $J_y$ & $\phi_2-\phi_1 =\pi/2\pm 2\pi n$ &  $\phi_1-\phi_2 =\pi/2\pm 2\pi n$ & $\phi_2-\phi_1 =\pi\pm 2\pi n$     \\ \hline
T\&R Amplitudes     & $|T_m|=1$, $|R_m|=0$                        & $|R_m|=1$, $|T_m|=0$   &  $|T_m|=|R_m|=1/\sqrt{2}$     \\ \hline
\end{tabular}
\caption{Relations between RIS types, T\&R coefficients and distribution of $J_y(\mathbf{r'})$, where $n=0,1,2,\cdots$.}
\label{tab:my-table1}
\end{table}

Note that the configuration shown in the above table are not the only way to achieve these three RIS types. In practice, the current may exhibit a continuous distribution within each element and have $x$ and $z$ components as well. As a result, the STAR-RIS elements can potentially achieve more complex and intricate phase-shift control. For example, the overall T\&R coefficients for an element may change with the angle of arrival or different polarization of the incident signal~\cite{liu2019generalized}. In this work, we only focus on scalar currents, i.e., $\mathbf{J}(\mathbf{r}')=\hat{\mathbf{y}}J_y(\mathbf{r}')$.

\section{Fundamental Performance Limits of Transmitting/Reflecting-Only RIS-Aided Single-User Scenario}

\begin{figure}[t!]
    \begin{center}
        \includegraphics[scale=0.35]{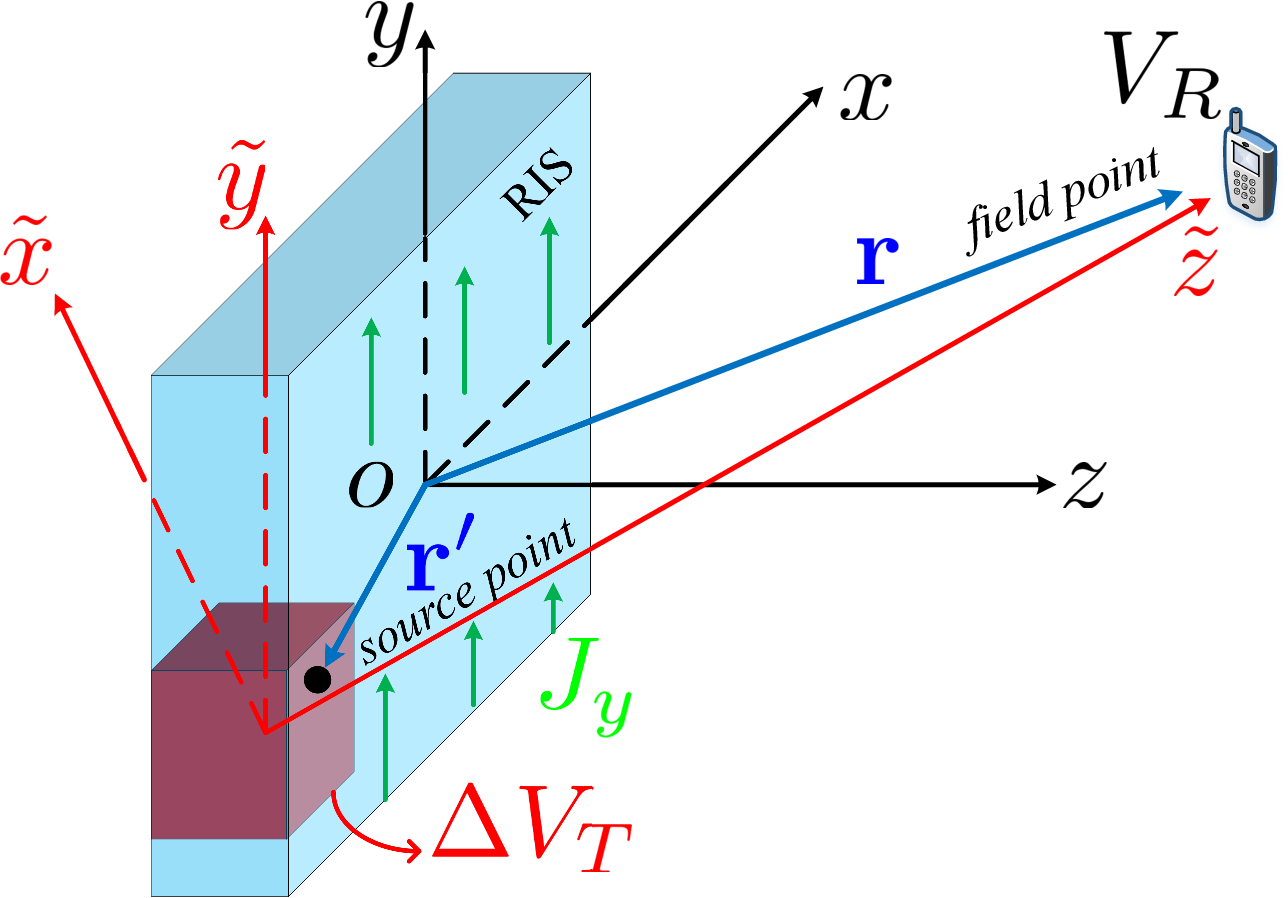}
        \caption{Geometrical illustration of transmitting/reflecting-only RIS and $\Delta V_T$.}
        \label{sysem_model}
    \end{center}
\end{figure}

Exploiting the proposed channel model, we first analyze the fundamental performance limits for transmitting-only and reflecting-only RISs, including the best achievable end-to-end channel gain, the maximum degrees of freedom of the end-to-end channel, and the power scaling law. The performance analysis of STAR-RIS-aided multi-user scenario is presented in the next section.
As illustrated in Fig.~\ref{sysem_model}, assume that the user is located in the $x$-$z$ plane. Thus, the kernel function defined in \eqref{K_def} can be calculated since the geometry of the problem is given. Exploiting \eqref{gyy}, the kernel function should take the following form:
\begin{equation}\label{krr}
    K(\mathbf{r}'_1,\mathbf{r}'_2) = \int_{V_R}\beta^2\frac{\exp{[-jk(|\mathbf{r}-\mathbf{r}'_2|-|\mathbf{r}-\mathbf{r}'_1|)]}}{16\pi^2|\mathbf{r}-\mathbf{r}'_1||\mathbf{r}-\mathbf{r}'_2|},
\end{equation}
where $\beta = j\omega\mu_0 + \frac{k^2}{j\omega\epsilon_0}$. 

\subsection{Field Boundary}
In previous studies, the field boundary is usually considered to be $r_b = \frac{2L^2}{\lambda}$~\cite{balanis2015antenna}, where $L$ is the largest dimension of the smart surface aperture. However, this result does not considered the angular position of the receiver, nor the physical size of the receiver. In our work, we investigate the field boundary exploiting the kernel function given in \eqref{krr}. For a receiver to be located within the far-field regime of the RIS, the phase-term in \eqref{krr} should not vary within the receiving volume $V_R$~\cite[eq. (88)]{miller2000communicating}. At the same time, the phase-term should also stay roughly constant for any two source point $\mathbf{r}'_1$ and $\mathbf{r}'_2$ within $V_T$. If we use half of a period as the criteria, we have the following:
\begin{equation}\label{phased}
     \frac{k\Delta x_{T,max}\Delta x_{R,max}}{r}  < \frac{\pi}{2}, \ \text{and} \  \frac{k\Delta y_{T,max}\Delta y_{R,max}}{r}  < \frac{\pi}{2},
\end{equation}
where $\Delta x_{T,max}$, $\Delta y_{T,max}$ are the largest dimensions of $V_T$ in the $x/y$ direction and $\Delta x_{R,max}$, $\Delta y_{R,max}$ are the largest dimensions of $V_R$. By multiplying these results, we have the following lemma:
\begin{lemma}\label{boundary}
The boundary between the radiating near-field and far-field, i.e., the maximum distance for near-field communication is given by:
\begin{equation}\label{rb}
    r_b = \sqrt{\frac{2\Delta x_{T,max}\Delta y_{T,max}}{\lambda}}\sqrt{\frac{2\Delta x_{R,max}\Delta y_{R,max}}{\lambda}},
\end{equation}
or equivalently, for a user with a distance of $r$ to fall within the far-field regime of the RIS, the maximum volume of the RIS is given as follows:
\begin{equation}\label{vmax}
    \Delta V_{T,max} = \frac{\lambda r}{2\Delta x_{R,max}}\frac{\lambda r}{2\Delta y_{R,max}}2\Delta z_{T,max},
\end{equation}
where $z_{T,max}$ is the maximum width of the RIS.
\begin{proof}
The proof is straightforward by multiplying the results in \eqref{phased} and exploiting $k = 2\pi/\lambda$, and $\Delta V_{T,max} =(2\Delta x_{T,max})(2\Delta y_{T,max})(2\Delta z_{T,max})$.
\end{proof}
\end{lemma}
\begin{remark}
According to \textbf{Lemma}~\ref{boundary}, the field boundary not only depends on the size of the RIS, but also on the carrier frequency and the aperture size of the receiver. The near-field region expands with the increase of carrier frequency and the size of the receiver. For example, if the receiving device is also a large intelligent surface (LIS), then the field boundary $r_b$ will be significantly larger so that near-field effect need to be considered. In Table~\ref{case}, we provide the value of $r_b$ for several typical communication scenarios.
\begin{table}[!h]
\centering
\small
\begin{tabular}{|c|c|c|c|c|c|}
\hline
  & $f_c$   & $\lambda_c$ & RIS dimension & Receiver dimension & Boundary ($r_b$) \\ \hline
1 & 1000 MHz & 0.3 m        & 0.5 m               & 0.1 m               & 0.33 m            \\ \hline
2 & 5 GHz (802.11a)   & 60 mm         & 0.5 m               & 0.1 m               & 1.67 m            \\ \hline
3 & 10 GHz   & 30 mm         & 0.5 m               & 0.1 m               & 3.33 m            \\ \hline
4 & 10 GHz   & 30 mm         & 0.5 m               & 0.5 m               & 16.67 m           \\ \hline
5 & 60GHz (802.11ay)  & 5 mm       & 0.5 m               & 0.1 m               & 20 m              \\ \hline
6 & $\sim$ 500 THz (Li-Fi)   & $\sim$ 600 nm         & 0.5 m               & 0.1 m               & $\sim 1.6\times 10^5$ m            \\ \hline
\end{tabular}
\caption{Position of the near-field/far-field boundary.}\label{case}
\end{table}

As can be observed from the above table, for wireless communication in higher frequency, the volume of usable near-field regime is significantly larger. 
Moreover, for visible light frequency, if the RIS dimension is still considered to be in the order of magnitude of meters, the position of the field boundary is farther than any practical communicating distances. This indicates that near-field effect should always be considered for visible light communication.
As will be discussed in later sections, the near-field channel has higher degrees of freedom compare to the far-field channel.
\end{remark}

\subsection{An Upper Bound of the End-to-End Channel Gain}
According to previous results discussed in \textbf{Remark 1}, to obtain the best available received power for the receiver $V_R$, we need to find the eigenfunction and eigenvalues of $K(\mathbf{r}'_1,\mathbf{r}'_2)$, i.e., to solve for $J_y(\mathbf{r}'_1)$ in the integral equation \eqref{eigen}.
In order to keep our derivations cleaner, we simply the kernel function. Recall in \eqref{K_def}, the kernel function measures the averaged received power (within $V_R$) for two source point. If $\mathbf{r}'_1$ and $\mathbf{r}'_2$ are two far away from each other, their corresponding kernel will tend to zero due to the fast oscillating phase terms in \eqref{krr}. As a result, in \eqref{A_2}, the integral can be separately evaluated within smaller volumes of sizes $\Delta V_{T}$ whose size is given in \eqref{vmax}.
Since we are now evaluating the integral in smaller volumes, we can assume that each volume of $\Delta V_T$ and $V_R$ are far apart compared with their sizes. As a result, exploiting the paraxial approximation, the kernel function can be expressed as follows~\cite{miller2000communicating}:
\begin{align}\label{krr_2}
    K(\mathbf{r}'_1,\mathbf{r}'_2) &= \beta^2\int_{V_R}\frac{1}{(4\pi r)^2}F(\mathbf{r}'_1)F^*(\mathbf{r}'_2)\exp{\left[-\frac{jk\Tilde{x}(\Tilde{x}'_1-\Tilde{x}'_2)}{r} -\frac{jk\Tilde{y}(\Tilde{y}'_1-\Tilde{y}'_2)}{r} \right]}\mathrm{d}V_R,
\end{align}
where 
\begin{equation}\label{focus}
    F(\mathbf{r}') = \exp{\left\{-jk\left[\Tilde{z}'-\frac{1}{2r}\left((\Tilde{x}')^2+(\Tilde{y}')^2 \right) \right] \right\}}
\end{equation}
is the focusing phase function within $V_T$ and $r$ is the distance between the chosen small volume of $\Delta V_T$ and $V_R$, $\Tilde{x}'$, $\Tilde{y}'$, and $\Tilde{z}'$ are the coordinates of the source point $\mathbf{r}'$ in the coordinate originated at the center of $\Delta V_T$ (as illustrated in Fig.~\ref{sysem_model}).
For the above approximation to be valid, the phase terms in \eqref{krr_2} should not osculate within the volume of $V_R$. 
Based on \textbf{lemma}~\ref{boundary}, this is guaranteed if we take $\Delta V_T\approx \frac{\lambda r}{2\Delta x_{max}}\frac{\lambda r}{2\Delta y_{max}}2\Delta z_T$.
We denote $N = \frac{V_T}{\Delta V_T}$ as the number of smaller volumes which $V_T$ is divided into.
Thus, we can evaluate the integral separately within each volume of $\Delta V_T$, i.e.,
\begin{equation}\label{inte}
    \int_{V_T} K(\mathbf{r}'_1,\mathbf{r}'_2)J_y(\mathbf{r}'_2)\mathrm{d^3}\mathbf{r}'_2 = \sum_{i=1}^N \int_{(\Delta V_T)_i} K_i(\mathbf{r}'_1,\mathbf{r}'_2)J_y(\mathbf{r}'_2)\mathrm{d}(\Delta V_T)_i,
\end{equation}
where $i\in[1,N]$ is the index of each volume $\Delta V_T$, $r_i$ is the distance between $(\Delta V_T)_i$ and $V_R$, and
\begin{equation}\label{ki}
    K_i(\mathbf{r}'_1,\mathbf{r}'_2) = \beta^2F(\mathbf{r}'_1)F^*(\mathbf{r}'_2)\frac{V_R}{(4\pi r_i)^2}
\end{equation}
is the value of the kernel function within $(\Delta V_T)_i$.
%%%%%%%%%
To obtain the upper bound of the channel gain, we assume that $J_y(\mathbf{r}')$ can be configured to a detailed distribution within $\Delta V_T$. According to \eqref{krr_2}, the following current source distribution yields the largest eigenvalue (see also \cite[Eq. (92)]{miller2000communicating}):
\begin{equation}\label{jy}
    J_y(\mathbf{r}')_i = \begin{cases}
       F_i(\mathbf{r}')/\sqrt{\Delta V_T}, & \text{if}\ \mathbf{r'} \in (\Delta V_T)_i \\
      0, & \text{otherwise},
    \end{cases}
\end{equation}
where $F_i(\mathbf{r}')$ is the focusing function given in \eqref{focus}. Note that for $F_i(\mathbf{r}')$, the coordinate $\Tilde{x}-\Tilde{y}-\Tilde{z}$ used in \eqref{focus} is located at the center of $(\Delta V_T)_i$. The focusing function ensures that the phase terms of all source point $\mathbf{r}'$ within $(\Delta V_T)_i$ add up constructively at $V_R$. Thus, by substituting \eqref{jy} into \eqref{inte}, we have the following theorem.
\begin{theorem}
The maximum channel gain between the RIS and a near-field receiver takes the following form:
\begin{equation}\label{A_result}
    |h_{\text{end-to-end}}|^2 =\begin{cases}
       \frac{DA_T}{4\pi d^2}\left(j\omega\mu_0 + \frac{k^2}{j\omega\epsilon_0}\right)^2\cdot\sum_{i=1}^N\frac{V_R\Delta V_T}{(4\pi r_i)^2}, & \text{if}\ V_T > \Delta V_T \\
      \frac{DA_T}{4\pi d^2}\left(j\omega\mu_0 + \frac{k^2}{j\omega\epsilon_0}\right)^2\cdot\frac{V_R V_T}{(4\pi r)^2}, & \text{if}\ V_T \leq \Delta V_T.
    \end{cases}
\end{equation}
\begin{proof}
Substituting \eqref{jy} into \eqref{inte} and \eqref{A_2}, we have the following:
\begin{equation}
    |h_{\text{end-to-end}}|^2 = \frac{DA_T}{4\pi d^2}\int_{V_T} J^*_y(\mathbf{r}'_1) \sum_{i}^{N} \left[\int_{(\Delta V_T)_i} K(\mathbf{r}'_1,\mathbf{r}'_2)\frac{F_i(\mathbf{r}'_2)}{\sqrt{\Delta V_T}}\mathrm{d}(\Delta V_T)_i\right] \mathrm{d^3}\mathbf{r}'_1
\end{equation}
Note that the integration within $V_T$ can be split into $N$-sum since $J_y(\mathbf{r}'_i)$ is zero outside $(\Delta V_T)_i$. Thus, the other integral can be evaluated in a similar way where the focusing function in $J_y$ cancels out the phase terms in the kernel function:
\begin{equation}
    |h_{\text{end-to-end}}|^2 =  \frac{DA_T}{4\pi d^2}\int_{V_T} \frac{F^*(\mathbf{r}'_1)}{\sqrt{\Delta V_T}}  \sum_{i} \left[\beta^2 F_i(\mathbf{r}'_1) \frac{V_R}{(4\pi r_i)^2}\sqrt{\Delta V_T}\right] =  \frac{A_T}{4\pi d^2}\beta^2\sum_{i=1}^N\frac{V_R\Delta V_T}{(4\pi r_i)^2}.
\end{equation}
Finally, by using the definition of $\beta$ in \eqref{krr}, the theorem is proved.
\end{proof}
\end{theorem}

% \begin{remark}
% In \eqref{A_result}, one cannot use volume $\Delta V_T$ larger than what is given in \eqref{vmax} to calculate the channel gain. However, one can use a smaller volume, for example, the volume of each STAR-RIS element if it is smaller than $\Delta V_{T, max}$.
% \end{remark}

\subsection{Maximum Degrees of Freedom of the End-to-End Channel}
According to \eqref{A_result}, the maximum channel gain is a summation of $N$ terms. In the following corollary, we prove that these $N$ components are in fact parallel channels. This demonstrates an important difference between near-field and far-field RIS-aided communication. Within the far-field regime, the channel is a rank-one matrix for free-space transmission. However, as we will show in the next corollary, due to the varying path lengths between different parts of the RIS and the receiver, the near-field end-to-end channel potentially has higher degrees of freedom.
\begin{corollary}
For RIS-aided near-field communication, the available degrees of freedom for multiplexing is:
\begin{equation}\label{dof}
    N = \begin{cases}
       1, & \text{if}\ V_T \leq \Delta V_T \\
       \frac{2V_TV_R}{(\lambda r)^2 \Delta z_T \Delta z_R}, & \text{otherwise},
    \end{cases}
\end{equation}
where $r$ is the distance between the centers of RIS and the receiving volume, $z_T$ and $z_R$ are the maximum width of the RIS and the receiver, respectively.
\begin{proof}
For convenience, we denote the current distribution within $(\Delta V_T)_i$ as $J_y^{(i)}(\mathbf{r}')$ and its generated E-field within $V_R$ as $E_y^{(i)}(\mathbf{r})$.
We only need to prove that $E_y^{(i)}(\mathbf{r})$ are orthogonal to each other within $V_R$. Note that two  functions are orthogonal if their inner product is zero, i.e., $<\mathbf{E}_1,\mathbf{E}_2> = \int_{V_R} \mathbf{E}^T_1\mathbf{E}^*_2\ dV_R=0$, where $\mathbf{E}^T$ is the transpose of vector $\mathbf{E}$ and $^*$ denotes the complex conjugate. Thus, we have $<E_y^{(p)}(\mathbf{r}),E_y^{(q)}(\mathbf{r})>=<J_y^{(p)},\mathcal{K}J_y^{(q)}>$, where $\mathcal{K}$ is the operator of the kernel function $K$, i.e., $\mathcal{K}J_y(\mathbf{r}') = \int_{V_T}K\cdot J_y(\mathbf{r}')dV_T$. According to \eqref{jy}, $<J_y^{(p)},J_y^{(q)}>=0$ if $p\neq q$ and $J_y^{(q)}$ is the eigenfunction of $\mathcal{K}$. As a result, we have $<E_y^{(p)}(\mathbf{r}),E_y^{(q)}(\mathbf{r})>=0$ for $p\neq q$. This is to say, the number of orthogonal equivalent source functions is equal to the number of volumes $\Delta V_T$, i.e., $N = V_T/(\Delta V_T)$. Finally, by exploiting the expression for $\Delta V_T$ in \eqref{vmax}, the corollary is proved.
\end{proof}
\end{corollary}

\begin{remark}
According to \eqref{dof}, for the case where the aperture size of the RIS is larger and the carrier frequency is higher, the available DoF is higher. However, utilize these DoF is a challenge due to the lack of RF chains on RISs. In theory, it is possible to send a superimposed signal to the RIS and these parallel signals excited different surface current patterns, and thus, at the receiver, the signals can be independently decoded by summing up the signals with weights according to the eigenfunctions within the $V_R$~\cite[Fig.~2]{9139337}.
\end{remark}
\subsection{Power Scaling Law}
Next, we present the near-field and far-field power scaling laws. In \eqref{jy}, we used normalized currents within $V_T$, i.e., $\int_{V_T} J_y^*J_y \mathrm{d}V_T = 1$. However, in practice, if we consider the RIS is illuminated by a far-field base station, we need to use the unnormalized currents\footnote{The unnormalized current is practically-relevant since a larger RIS collects more incident energy. In Theorem 1, however, the normalized current is used since we already involve the aperture area of STAR-RIS ($A_T$) in the equation.}, i.e., $J_y(\mathbf{r}') = F(\mathbf{r}')$. Thus, we have the following corollary: 
\begin{corollary}
Suppose that the RIS has $M$ elements, the power scaling law is given as follows:
\begin{equation}\label{scaling}
    P_r \propto V_R \sum_{i}^{N}\frac{V_{ele}^2}{(4\pi r_i)^2}M^2_i,
\end{equation}
where $P_r$ is the received power, $V_{ele}$ is the volume of each RIS element and $M_i$ is the number of element within $(\Delta V_T)_i$, such that $\sum_{i=1}^N M_i = M$.
\begin{proof}
The corollary is proved by substituting $J_y(\mathbf{r}') = F(\mathbf{r}')$ into \eqref{A_2} and following the similar derivations in \textbf{Theorem 1}.
\end{proof}
\end{corollary}

\begin{remark}\label{vt_meaning}
Far away from the field boundary, the power scaling law in \eqref{scaling} can be further simplified. Within the far-field regime, the volume $\Delta V_{T,max}$ given in \eqref{vmax} is larger than the entire volume of RIS. Thus, the summation can be simply removed and we arrive at the similar far-field power scaling law as given in previous works:
\begin{equation}\label{scaling_far}
    P^{\text{far-field}}_r \propto V_R \left(\frac{V_{ele}}{4\pi \bar{r}}\right)^2\cdot M^2,
\end{equation}
Within the near-field region when the volume $\Delta V_{T,max}$ is smaller or equal to the size of $V_{ele}$, there are only one element within each $(\Delta V_T)_i$ and N = M. Thus, the received power starts to scale linearly with $M$:
\begin{equation}\label{scaling_near}
    P^{\text{near-field}}_r \propto V_R \sum_{m=1}^{M} \left(\frac{V_{ele}}{4\pi r_m}\right)^2 \approx V_R \left(\frac{V_{ele}}{4\pi \bar{r}}\right)^2\cdot M,
\end{equation}
where $\bar{r}$ is the distance between the centers of the RIS and the receiver.
\end{remark}

\begin{table}[!t]
\small
\begin{center}
\begin{tabular}{|c|c|c|c|c|}
\hline
           & Field region & End-to-end channel gain                                  & Degrees of freedom ($N$) & Power scaling law \\ \hline
Far-field  & $r>r_b$      & $\propto \frac{V_RV_T}{(4\pi r)^2}$ & $1$                &  $\propto M^2$                 \\ \hline
Near-field & $r_r<r\leq r_b$  & $\propto \sum_{i}^N\frac{V_R\Delta V_T}{(4\pi r_i)^2}$ & $\frac{2V_TV_R}{(\lambda r)^2 \Delta z_T \Delta z_R}$                &    $\propto \sum_i^N (M_i^2/r^2_i)$                 \\ \hline
\end{tabular}
\caption{Comparing fundamental performance limits of the radiating near-field and far-field regimes,}\label{tab:my-table}
\end{center}
where $r_r$ indicates the boundary of the reactive near-field. According to antenna theory, this value is typically given by $r_r = 0.62\sqrt{L^3/\lambda}$, where $L$ is the largest dimension of the RIS.
\end{table}
\subsection{Summary}
For the convenience of the readers, we summarize and compare the performance limits for near-field and far-field regimes as derived in the above subsections in Table~\ref{tab:my-table}.

\section{Performance Analysis for the STAR-RIS-Aided Multi-User Scenario}

\begin{figure}[b!]
    \begin{center}
        \includegraphics[scale=0.35]{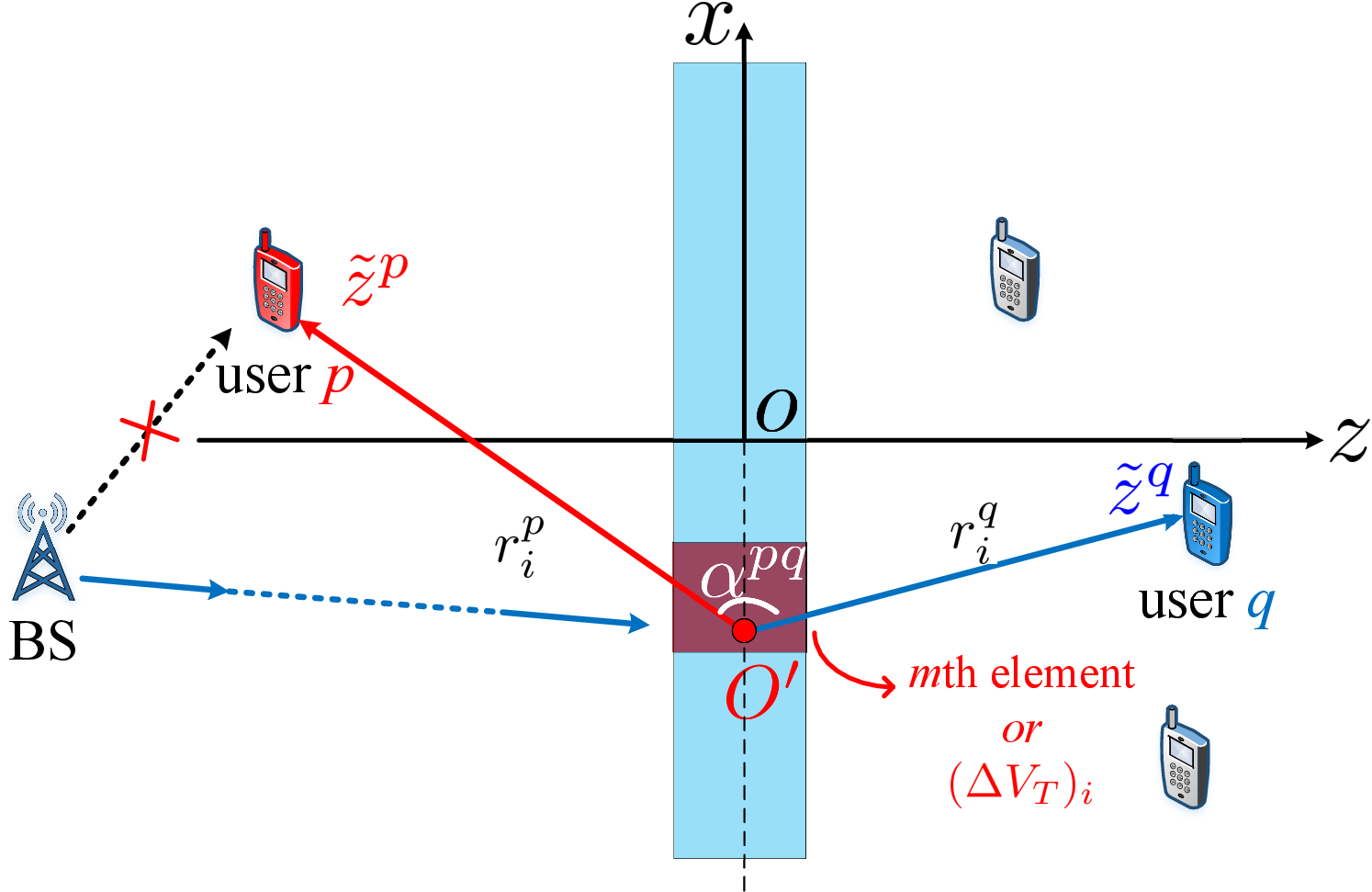}
        \caption{Illustration for the STAR-RIS-aided multi-user scenario.}\label{fig_two}
    \end{center}
\end{figure}

In this section, we extend our analysis to the case where multiple users are simultaneously served by the STAR-RIS. We first consider the case where the multiple users are all within the near-field regime and propose three configuration strategies. Then, we consider the hybrid near-field and far-field case where the transmission-side (indoor) user is within the near-field and the reflection-side (outdoor) user is within the far-field.
As illustrated in Fig.~\ref{fig_two}, assume that the total number of user is $U$. To demonstrate the \textit{STAR} functionality, we assume that the users are located on both sides of the STAR-RIS.
According to our proposed channel model, for the multi-user case, the channel gain for user $p$ is given as follows:
\begin{equation}\label{hab}
    |h^{(p)}|^2 =  \frac{DA_T}{4\pi d^2}\int_{V_T} J^*_y(\mathbf{r}'_1) \int_{V_T} K^{(p)}(\mathbf{r}'_1,\mathbf{r}'_2)J_y(\mathbf{r}'_2)\mathrm{d^3}\mathbf{r}'_1 \mathrm{d^3}\mathbf{r}'_2,
\end{equation}
where $K^{(p)}(\mathbf{r}'_1,\mathbf{r}'_2)$ is the kernel function corresponding to user $p$.
Similar to the single-user case, we evaluate the above integral within each smaller volumes of sizes $\Delta V_T$. However, the difference is that for the multi-user case, $J_y(\mathbf{r}'_2)$ need to be optimized for all users. In the following, we give the fundamental performance limit of STAR-RIS serving multiple near-field users in terms of the sum of their channel gains.

To start with, we exploit the fact that only the term $K^{(p)}(\mathbf{r}'_1,\mathbf{r}'_2)$ in \eqref{hab} is related to the users position. Thus, the sum of the channel gains for all $U$ users can be expressed as follows:
\begin{equation}\label{sum_h2}
     \sum_p^U |h^{(p)}|^2 =  \frac{DA_T}{4\pi d^2}\int_{V_T} J^*_y(\mathbf{r}'_1) \int_{V_T} \left(\sum_p^U  K^{(p)}(\mathbf{r}'_1,\mathbf{r}'_2)\right )J_y(\mathbf{r}'_2)\mathrm{d^3}\mathbf{r}'_1 \mathrm{d^3}\mathbf{r}'_2,
\end{equation}
where $\sum_p^U  K^{(p)}(\mathbf{r}'_1,\mathbf{r}'_2)$ is the sum of the kernel function for all users.
Here, we assume that the STAR-RIS elements are smaller than $V_{T,max}$. According to last section, the value of the kernel function for user $p$ within the $m$th element $(V_m)$ is given in \eqref{ki} as $ K^{(p)}_i(\mathbf{r}'_1,\mathbf{r}'_2) = \beta^2F_i(\mathbf{r}'_1)F_i^*(\mathbf{r}'_2)\frac{V_R}{(4\pi r^{(p)}_i)^2}$, where the focus function is $F_i(\mathbf{r}') =  \exp{\left\{-jk\left[\Tilde{z}_i'-\frac{1}{2r}\left((\Tilde{x}_i')^2+(\Tilde{y}_i')^2 \right) \right] \right\}}$. Note that the $\Tilde{x}_i'$, $\Tilde{y}_i'$, and $\Tilde{z}_i'$ are evaluated in the \textit{local} coordinate centered at $(\Delta V_T)_i$ where its $z_i$-axis points in the direction of user $p$. Thus, calculating the sum of the kernel functions involves transformations between all these \textit{local} coordinates:

\begin{align}
    \sum_p^U K^{(p)}(\mathbf{r}'_1,\mathbf{r}'_2) &= \beta \cdot V_R  \sum_p^U \sum_i^{N}\frac{1}{(4\pi r_i^{p})^2} F^{(p)}_i(\mathbf{r}'_1)F^{(p)*}_i(\mathbf{r}'_2),\\ \label{aff}
    &=\frac{\beta \cdot V_R}{(4\pi)^2}  \sum_i^N \left( \sum_p^U \frac{F^{(p)}_i(\mathbf{r}'_1)}{(r^p_i)^2}F^{(p)*}_i(\mathbf{r}'_2) \right),
\end{align}
where $F^{(p)}_i(\mathbf{r}')$ is the focus function for $(\Delta V_T)_i$ and user $p$.
Unlike the single-user case in \eqref{ki}, variables $\mathbf{r}'_1$ and $\mathbf{r}'_2$ in \eqref{aff} are not separable in the summed kernel functions. As a result, there are no trivial solutions for $\mathbf{J}_y(\mathbf{r}')$ which could maximize the summed channel gain in \eqref{sum_h2}.
% The current configuration which maximums the summed channel gain for all users can only be obtained through the following optimization problem:
% \begin{equation}
% \begin{aligned}
% \max_{\mathbf{J}_y(\mathbf{r}')_i} \quad & \int_{V_T} J^*_y(\mathbf{r}'_1)_i \int_{V_T} \left( \sum_p^U \frac{F^{(p)}_i(\mathbf{r}'_1)}{(r^p_i)^2}F^{(p)*}_i(\mathbf{r}'_2) \right)J_y(\mathbf{r}'_2)_i\mathrm{d^3}\mathbf{r}'_1 \mathrm{d^3}\mathbf{r}'_2,\\
% \textrm{s.t.} \quad & \int_{(\Delta V_T)_i}{\mathbf{J}_y(\mathbf{r}')_i}{\mathbf{J}^*_y(\mathbf{r}')_i} = 1,\\
% \end{aligned}
% \end{equation}
% where ${\mathbf{J}_y(\mathbf{r}')_i}$ is the current configuration within $(\Delta V_T)_i$ and the constraint guarantees that the equivalent current is normalized.
\begin{figure*}[t!]
\centering
\subfigure[PS]{\label{n01}
\includegraphics[width= 1.3in]{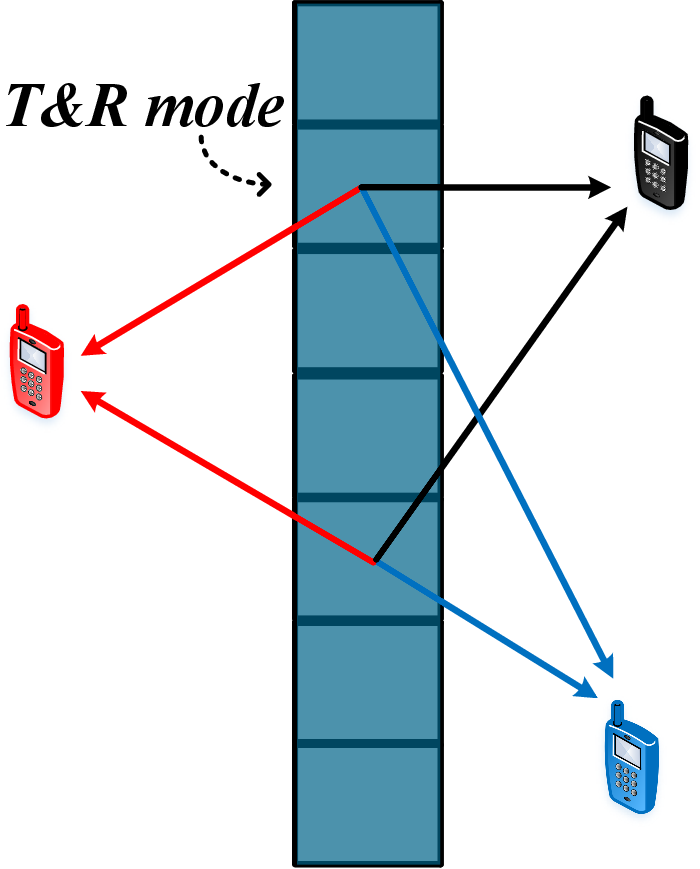}}
\subfigure[SEG]{\label{nb1}
\includegraphics[width= 1.3in]{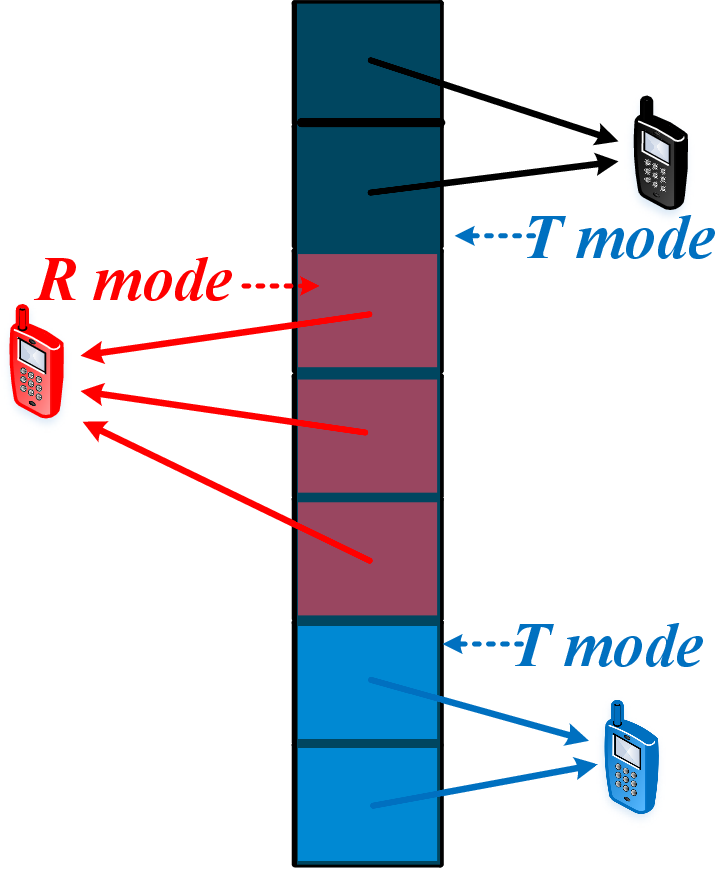}}
\subfigure[REG]{\label{na1}
\includegraphics[width= 1.3in]{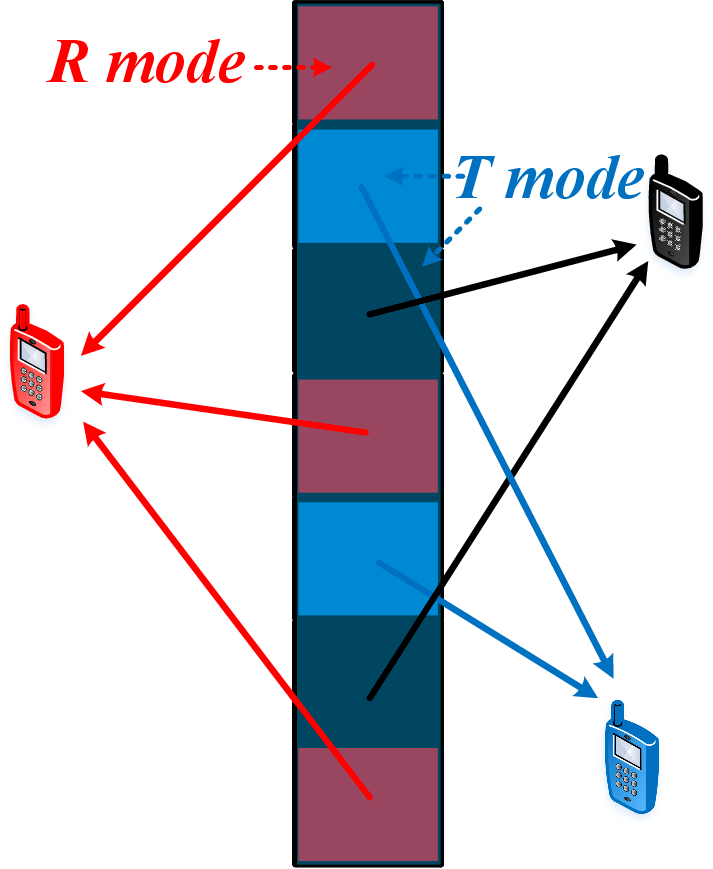}}
\caption{Conceptual illustration of STAR-RIS configuration strategies, where \textit{T mode} refers to full transmission mode, \textit{T mode} refers to full reflection mode, and \textit{T\&R mode} refers to simultaneous transmission and reflection mode~\cite{mu2021simultaneously}.}\label{nice1}
\end{figure*}
In practice, the channel and positional information of users are difficult to obtain in real time.
As a result, we propose three heuristic strategies for configuring $\mathbf{J}_y(\mathbf{r}')$ in the multi-user scenario, namely the power splitting (PS), the selective element grouping (SEG), and the random element grouping (REG) strategies.

As illustrated in Fig.~\ref{nice1}, for the PS strategy, all the STAR-RIS elements split their power equally to simultaneously serve all users. For the REG and SEG strategies, elements are divided into $U$ groups so that each group has roughly $M/U$ elements. Elements in each group are dedicated to serve one single user. In REG, the element grouping is randomly assigned whereas in SEG, we select $M/U$ elements which are closest to the user on the STAR-RIS. In the following, we discuss the current distribution configuration and performance of these three strategies in detail.

\subsection{Power Splitting (PS) Strategy}
As illustrate in Fig.~\ref{n01}, in PS strategy, all STAR-RIS elements simultaneously serve all users. To achieve that, for the $m$th element, the current is configured as the linear superposition of the focusing functions for all users, normalized by the volume of the element ($V_m$), i.e., 
\begin{equation}\label{jy2}
    J_{y,m}(\mathbf{r}') = 
       \sum_{p=1}^U F^{(p)}_m(\mathbf{r}')/\left(\sqrt{V_mA_m} \right),
\end{equation}
where $F^{(p)}_m(\mathbf{r}')$ is the focus function for user $p$ within the $m$th element (defined in \eqref{focus}) and ${V_m}A_m = \int_{V_m}||\sum_{p=1}^U F^{(p)}_m(\mathbf{r}')|| \mathrm{d} V_m$ is the normalization factor. 
Next, we evaluate the channel gain for the users. To obtain tractable results, we further assume that the size of each STAR-RIS element is small compared to $r^p_i$. Thus, the kernel function for the $m$th element can be reduced to $K_m^{(p)}(\mathbf{r}'_1,\mathbf{r}'_2) \approx \frac{\beta^2V_R}{(4\pi r^p_m)^2}e^{-jk\Tilde{z}^{p}_1}e^{+jk\Tilde{z}^{p}_2}$, where $\Tilde{z}^{p}_{1/2}$ are the $z$-coordinate of the source point $\mathbf{r}'_{1/2}$ in their own orientates originated at the center of the $m$th element and $r_m^p$ is the distance between the $m$th element and user $p$. With this simplification, the end-to-end channel gain can be formulated in a closed-form expression.
\begin{theorem}\label{t_ps}
For the PS strategy, the end-to-end channel gains for user $q$ is given as follows:
\begin{equation}\label{ha2}
    |h^{(q)}|^2 = \frac{DA_T\beta^2}{4\pi d^2}\sum_{m=1}^{M} \frac{V_RV_m}{(4\pi r^q_m)^2 A_m}\left(1+ \sum_{p\neq q}\mathrm{sinc}(\xi^{pq}_m) \right)^2,
\end{equation}
where where $\beta = j\omega\mu_0 + \frac{k^2}{j\omega\epsilon_0}$ is the constant defined earlier, $\xi^{pq}_m = \pi\cdot(1-\cos\alpha^{pq}_m)\Delta z_m/\lambda$, $\Delta z_m$ is the width of each STAR-RIS element, and $\alpha^{pq}_m$ is the angle between $O'p$ and $O'q$, for $O'$ being the center of the $m$th element, as illustrated in Fig.~\ref{fig_two}.
\begin{proof}
See Appendix~\ref{ap_c}.
\end{proof}
\end{theorem}

\subsection{Random Element Grouping (REG) Strategy}
As illustrate in Fig.~\ref{na1}, in REG strategy, the STAR-RIS elements are randomly assigned to focus a single user. Similar to the last subsection, we assume that the STAR-RIS elements are small so that within the volume of $\Delta V_T$, there are multiple elements. Thus, the current configuration for REG strategy is given as follows:
\begin{equation}\label{jy3}
    J_{y,m}(\mathbf{r}')_i = \begin{cases}
      F^{(p)}_m(\mathbf{r}')/\sqrt{V_m}, & \text{if}\ \mathbf{r'} \in V_m\ \text{and}\ m \in \mathcal{M}^p\\
      0, & \text{otherwise},
    \end{cases}
\end{equation}
where $F^{(p)}_m$ is the focus function for user $p$ within $V_m$ defined in \eqref{focus} and $\mathcal{M}^p$ denotes the group of elements chosen to focus user $p$. For this configuration, we show the channel gain in the following theorem.

\begin{theorem}\label{t_reg}
For the REG strategy, the end-to-end channel gains for user $p$ are given as follows:
\begin{equation}\label{h_reg}
    |h^{(p)}|^2 = \frac{DA_T\beta^2}{4\pi d^2}\sum_{i=1}^{N^{(p)}} \frac{V_RV_m}{(4\pi r^p_i)^2}\left( M^{(p)}_i +  \sum_{q\neq p} M^{(q)}_i\mathrm{sinc}(\xi^{pq}_i) \right)^2,
\end{equation}
where $r_i^{(p)}$ is the distance between the center of $(\Delta V_T)_i$ and user $p$, $V_m$ is the volume of each STAR-RIS element, $M^{(p)}_i$ is the number of elements in $(\Delta V_T)_i$ whose phase shifts are optimized for user $p$, $N^{(p)} = \frac{2V_TV_R}{(\lambda r^p)^2 \Delta z_T \Delta z_R}$ is the achievable degrees of freedom for user $p$,  $\xi^{pq}_i = \pi\cdot(1-\cos\alpha^{pq}_i)\Delta z_m/\lambda$, $\Delta z_m$ is the width of each STAR-RIS element, and $\alpha^{pq}_i$ is the angle between $O'p$ and $O'q$, for $O'$ being the center of $(\Delta V_T)_i$, as illustrated in Fig.~\ref{fig_two}.
\begin{proof}
See Appendix~\ref{ap_d}.
\end{proof}
\end{theorem}

\begin{remark}
The channel gain for user $p$ should reduces to the single-user case if all elements are configured for this user. This can be easily verified by letting $M_i^{(q)} = 0$ in \eqref{ha2}. In other cases, the channel gain for a given user is always smaller compared to the upper bound given in \eqref{A_result}. This gap is determined by the argument of $\xi^{pq}_i$, which itself depends on the locations of the users.
\end{remark}

\subsection{Selective Element Grouping (SEG) Strategy}
The path lengths between the near-field user and different STAR-RIS elements varies significantly. Thus, unlike in the far-field region, element grouping has a greater effect on the overall channel gains for all users. As illustrate in Fig.~\ref{nb1}, for the SEG strategy, the elements which are closest to one user are all assigned to this particular user. In this case, the STAR-RIS can be regrouped into $U$ volumes, the $p$th volume contains all elements assigned to user $p$. The channel gain can be calculated as follows:
\begin{corollary}
For the SEG strategy, the end-to-end channel gains for user $p$ are given as follows:
\begin{equation}\label{h_SEG}
    |h^{(p)}|^2 = \frac{DA_T\beta^2}{4\pi d^2} \left( \sum_{i\in \mathcal{V}^p}\frac{V_RV_m}{(4\pi r^p_i)^2}M_i^2 +  \sum_{i'\in \mathcal{V}^q, q\neq p}\frac{V_RV_m}{(4\pi r^p_{i'})^2} \left(M_{i'} \mathrm{sinc}(\xi^{pq}_{i'})\right)^2 \right) ,
\end{equation}
where $\mathcal{V}^p$ denotes the volume of STAR-RIS which is assigned to user $p$, the index $i$ denotes the volume $(\Delta V_T)_i$, $r^{p}_i$ is the distance between $(\Delta V_T)_i$ user $p$, $M_i$ is the number of elements within $(\Delta V_T)_i$.
\begin{proof}
Since SEG is a special case of REG strategy, \eqref{h_SEG} can be derived from \eqref{h_reg} by using the fact all element within $(\Delta V_T)_i$, $i\in \mathcal{V}^p$ is assigned to user $p$ and thus $M^{q}_i = 0$ if $q \neq p$.
\end{proof}
\end{corollary}

\begin{remark}\label{eg}
The performance of SEG strategy is strictly better than that of the REG strategy. Comparing the the results in \eqref{h_reg} and \eqref{h_SEG}, we have $r^p \leq r^p_i, \ \forall i$. Thus, for each element, the coefficients in the SEG is larger than those in the REG strategy. This difference is more significant within the near-field regime due to the greater differences between path-lengths $r^p_i$. However, to obtain the element grouping configuration for SEG, positional information for all users is required which may significantly increase the overhead of the channel estimation procedure.
\end{remark}
\subsection{Performance Analysis in the Hybrid Near-Field and Far-Field Regimes}\label{sec_cs}
\begin{figure}[t!]
    \begin{center}
        \includegraphics[scale=0.4]{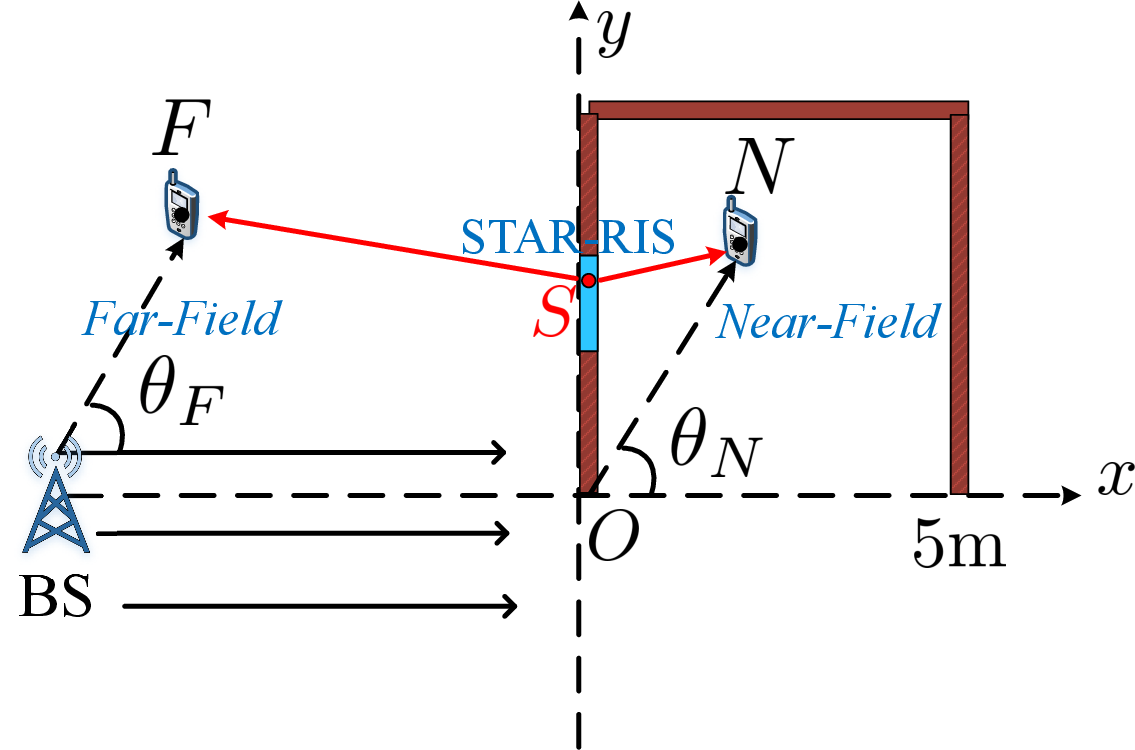}
        \caption{Geometrical setup for the considered hybrid near-field and far-field scenario.}
        \label{system_case}
    \end{center}
\end{figure}
\begin{table}[!b]
\centering
\small
\begin{tabular}{|c|c|c|c|c|c|c|c|}
\hline
$f_c$   & $\lambda_c$ & $z_{\text{STAR}}$ & $L_{\text{STAR}}$ & $A_{F/N}$ & $r_b$ & $r_{SF}$ & $r_{SN}$ \\ \hline
$30$GHz & $9.9$mm     & $0.05$m & 0.5m              & 0.01m$^2$ & 10m & $2$m & $20$m \\ \hline
\end{tabular}
\caption{Parameters for the considered hybrid near-field and far-field scenario.}
\label{para}
\end{table}
In the above sections, we investigated the performance of the PS, SEG, and REG strategies within the near-field regime. However, in practical applications, users could be located in both the near-field and the far-field regimes.
Here, we evaluate the performance of STAR-RISs in the hybrid near-field and far-field regimes. Specifically, we investigate an outdoor-to-indoor communication scenario where the STAR-RIS is deployed on the window to simultaneously transmit and reflect wireless signals to indoor and outdoor users, respectively.
As illustrated in Fig.~\ref{system_case}, we consider the case where mobile users are located both indoor and outdoor. The origin $O$ is at bottom left of the room. The BS is far from the origin and its main lobe is pointing in the $x$-direction. Thus, the wireless signal can be regard as a plane wave near the origin. As shown in the figure, the wireless signal cannot penetrate walls and can only enters through the opened bottom edge of the room ($x$-axis) or through the STAR-RIS on the $y$-axis. For simplicity, we further assume that the system is translation invariant in the $z$-direction. In terms of multiple access, we assume that all users are served simultaneously. This can be achieved by employing frequency division multiple access. Furthermore, we assume that the base station antenna follows 3GPP model where the radiation pattern within each sector is as follows~\cite{8422746}:
\begin{equation}\label{D}
    D (\theta) = -\min \left\{12\left(\frac{\theta}{\theta_{\text{3dB}}}\right)^2, 30dB \right\},
\end{equation}
where $\theta_{\text{3dB}} = 65^\circ$ is the horizontal 3 dB beamwidth.

As illustrated in Fig.~\ref{system_case}, user $F$ is an outdoor, far-field receiver in the direction of $\theta_F$ with respect to the BS. User $N$ is an indoor user where the angle between $ON$ and the $x$-axis is $\theta_N$. In the following, we investigate how the end-to-end channel gain for the two users change with $\theta_F$ and $\theta_N$.

\subsubsection{Distribution of Signal Strength Without STAR-RISs}
First, we consider the benchmark scenario without STAR-RIS. Assuming the wall along the $y$-axis is completely sealed so that the signal cannot penetrate. Nevertheless, a small portion of the signal strength can still be received by user $F$ and $N$. This is because for user $N$, the wireless signal is diffracted at point $O$, and for user $F$, there are sidelobes of the BS antenna which radiates towards the horizontal direction of $\theta_2$. According to the radiation pattern given in \eqref{D}, the channel gain between BS and user $F$ can be expressed as follows:
\begin{equation}\label{hp}
    \left|h^F_{\text{end-to-end}}\right| \propto -\min \left\{12\left(\frac{\theta_F}{\theta_{\text{3dB}}}\right)^2, 30dB \right\}\cdot \frac{A_F}{r_{OF}},
\end{equation}
where $A_F$ is the aperture area of user $F$ and $r_{OF}$ is the length of $OF$.
Here, we assume that the contribution of the reflected signals (by the wall) at user P is weak and can be neglected.
For user $N$, the received signal power can also be expressed in terms of $\theta_N$ and distance $ON$. According to~\cite[eq. (39)]{booker1950concept}, the angular distribution of the signal power within the shadow edge (the first quadrant in Fig.~\ref{system_case}) can be expressed as follows\footnote{Here, the secondary reflections from the walls are neglected.}:
\begin{equation}\label{hq}
    \left|h^N_{\text{end-to-end}}\right| \propto  \frac{A_N}{k^2r_{ON}\lambda \sin^2\theta_N},
\end{equation}
where $A_N$ is the aperture area of user $N$ and $r_{ON}$ is the length of $ON$. It is shown in \eqref{hq} that the channel gain for user $N$ falls off inversely with $\sin^2\theta_N$ and the distance $r_{ON}$. This is consistent with common intuition that the top-left corner of the room should have the worst signal (see Fig.~\ref{n0} for the simulated result for the indoor radiation power distribution).

\subsubsection{Distribution of Signal Strength With the aid of STAR-RIS}
To improve the channel gain for both the outdoor and indoor users, we open a window in the wall on the $y$-axis and deploying a STAR-RIS. As illustrated in Fig.~\ref{system_case}, the STAR-RIS is plotted as a blue rectangle. Exploiting the criteria given in \eqref{rb}, we identify the field boundary for the STAR-RIS and user pair. According to the parameters in Table~\ref{para}, the field boundary is $r_b = 10$ m. This indicates that the indoor user $N$ always located within the near-field region and the outdoor user can be considered as far-field if the distance $r_{SF} > 10$ m. Next, we calculate the critical volume given in \textbf{Lemma}~\ref{boundary}:
\begin{equation}
    \Delta V_T^{F/N} = \frac{(\lambda r_{SF/SN})^2}{2A_{F/N}}z_{\text{STAR}} \approx 2.5 \times 10^{-4}\text{ m}\cdot r_{SF/SN}^2.
\end{equation}
For near-field user $N$, we have $r_{SN} = 2$ m, then we have $\Delta V_T^{N} \approx 0.02$ m$^2\cdot z_{\text{STAR}}$. This means that the degrees of freedom of the channel between STAR-RIS and user $N$ is $N^N = V_{T}/\Delta V_T \approx 12$. For far-field user $F$, we have $r_{SF} = 20$ m, then $\Delta V_T^{F} \approx 2$ m$^2\cdot z_{\text{STAR}}$. This volume is significantly larger than the STAR-RIS since the volume of STAR is only $V_{\text{STAR}} = 0.25$ m$^2\cdot z_{\text{STAR}}$.

To further obtain the channel gains for the indoor and outdoor users, we consider the employment of the three proposed STAR-RIS configuration strategies, i.e., PS, REG, and SEG strategies.
\begin{corollary}
For the PS strategy, the end-to-end channel gains between the BS and user P/Q through STAR-RIS are given as follows:
\begin{align}\label{hp_ps}
    |h^F_{PS}|^2 &= \frac{D_0A_T\beta^2}{4\pi d^2}\frac{V_RV_{STAR}(1+\text{sinc}(\xi^{FN}))^2}{(4\pi r^F)^2A_m},\\ \label{hq_ps}
    |h^N_{PS}|^2 &= \frac{D_0A_T\beta^2}{4\pi d^2}\sum_{m=1}^M \frac{V_RV_m}{(4\pi r^N_m)^2A_m}(1+\text{sinc}(\xi^{FN}_m))^2,
\end{align}
where $D_0 = D(\theta=0)$ is the directivity of the BS antennas, $A_T$ is the aperture size of STAR-RIS, $d$ is the distance between BS and STAR-RIS, $V_R$ is the volume of the receiver, $V_{STAR}$ is the volume of the STAR-RIS, $\xi^{FN} = \pi(1-\cos{\alpha^{FN}})\Delta z_m/\lambda$, $\alpha^{FN}$ is the angle between user $F$ and user $N$ observed at the center of STAR-RIS, and $V_m$ is the volume of each STAR-RIS element.
\begin{proof}
For the far-field user $F$, its distances and angles to different STAR-RIS elements are almost the same. Thus, by using $r^F_m = r^F$ and $\xi_m^{FN} = \xi^{FN}$ in \eqref{ha2}, \eqref{hp_ps} can be proved. For the near-field user $F$, its channel gain follows the same form in \eqref{ha2}.
\end{proof}
\end{corollary}

\begin{corollary}
For the REG and SEG strategy, the end-to-end channel gains between the BS and user P/Q through STAR-RIS are given as follows:
\begin{align}\label{hp_EG}
    |h^F_{REG}|^2 &= |h^F_{SEG}|^2 = \frac{D_0A_T\beta^2}{4\pi d^2}\frac{V_RV_m}{(4\pi r^F)^2}\left[(M^F)^2 + (M^N \text{sinc}(\xi^{FN}))^2 \right],\\ \label{hq_REG}
    |h^N_{REG}|^2 &= \frac{D_0A_T\beta^2}{4\pi d^2}\sum_{i=1}^{12} \frac{V_RV_m}{(4\pi r^N_i)^2}\left[M_i^N+ M_i^F\text{sinc}(\xi^{FN}_i)\right]^2,\\ \label{hq_SEG}
    |h^N_{SEG}|^2 &= \frac{D_0A_T\beta^2}{4\pi d^2}\left[\sum_{i\in \mathcal{V}^Q} \frac{V_RV_m}{(4\pi r^N_i)^2}M_i^2+ \sum_{i\in \mathcal{V}^F} \frac{V_RV_m}{(4\pi r^N_i)^2} \left(M_i\text{sinc}(\xi^{FN}_i)\right)^2\right],
\end{align}
where $\xi^{FN}_i = \pi(1-\cos{\alpha^{FN}_i})\Delta z_m/\lambda$, $\alpha^{FN}_i$ is the angle between user $F$ and user $N$ observed at $(\Delta V_T)_i$, and $M_i$ is the number of STAR-RIS element within $(\Delta V_T)_i$.
\begin{proof}
For the far-field user $F$, its distances and angles to different STAR-RIS elements can be regarded as the same. Thus, by using $r^F_m = r^F$ and $\xi_m^{FN} = \xi^{FN}$ in \eqref{h_reg}, \eqref{hp_EG} can be proved. For the near-field user $N$, its channel gain follows similar forms in \eqref{hq_REG} and \eqref{hq_SEG}.
\end{proof}
\end{corollary}
\begin{remark}
Comparing the channel gains for user $F$ and $N$ with and without the aid of STAR-RIS, it can be observed that the undesirable angle dependencies with $\theta_F$ and $\theta_N$ in \eqref{hp} and \eqref{hq} are removed in the STAR-RIS-aided links. This will be further illustrated in our numerical results (see Fig.~\ref{case_deg}).
\end{remark}

\begin{figure*}[b!]
\begin{center}
        \includegraphics[scale=0.6]{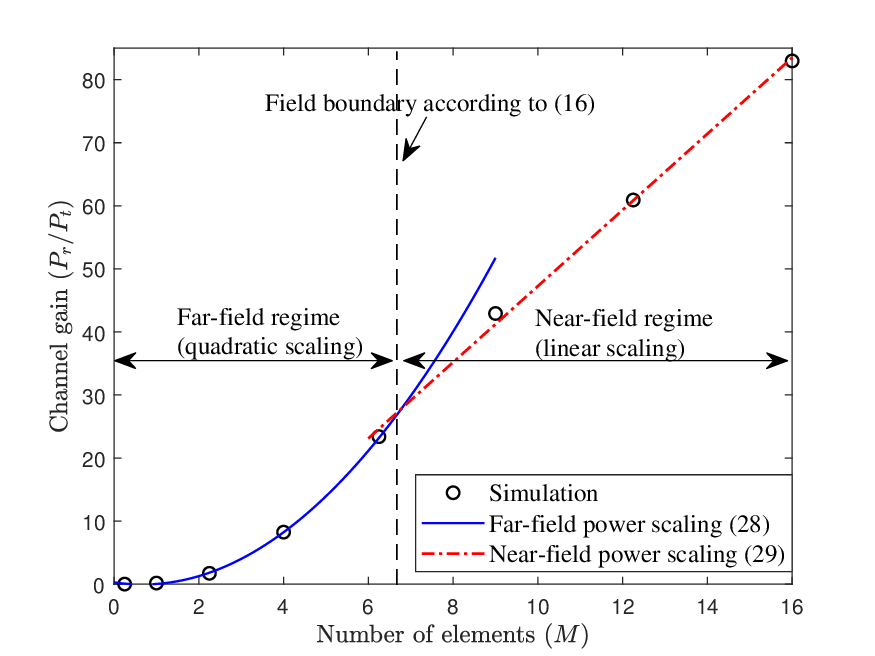}
       \caption{Near-field and far-field power scaling laws of transmitting/reflecting-only RIS.}\label{pscaling}
       \end{center}
\end{figure*}
\section{Numerical Results}

In this section, numerical results are provided to validate the proposed Green's function based channel model and power scaling laws for the transmitting/reflecting-only RIS-aided single user scenario. We also present simulation results for the three STAR-RIS configuration strategies and coverage simulation results for the STAR-RIS-aided multi-user scenario.

\subsection{Transmitting/Reflecting-Only RIS-Aided Single User Scenario}

\subsubsection{RIS Power Scaling Laws}
In Fig.~\ref{pscaling} we investigate the power scaling laws for the near-field and far-field regimes proposed in \eqref{scaling_near} and \eqref{scaling_far}, respectively. In the simulation, we assume that the receiver is located at a distance of $d=0.7$ m to the center of RIS and the dimension of the receiver is $z_R \approx 1.0$ cm. The number of RIS elements range from $1$ to $16$ and its size range from $1$ cm to $10$ cm. The wavelength of the carrier signal is $0.01$ m. As can be observed from the figure, since the position of the receiver is fixed, the user is within the far-field regime if the RIS has smaller size and the user is within the near-field regime if the RIS becomes larger. Within the far-field regime, the power scales with $M^2$, as derived in \eqref{scaling_far}. However, by increasing the size of the RIS, this quadratically growth cannot continue forever. As the size of the RIS reach the volume correspond to the boundary position, i.e., \eqref{vmax}, the user falls into the near-field regime. As indicated with the red dashed line, a linear power scaling law is observed, as derived in \eqref{scaling_near}.
\begin{figure*}[h!]
\begin{center}
        \includegraphics[scale=0.6]{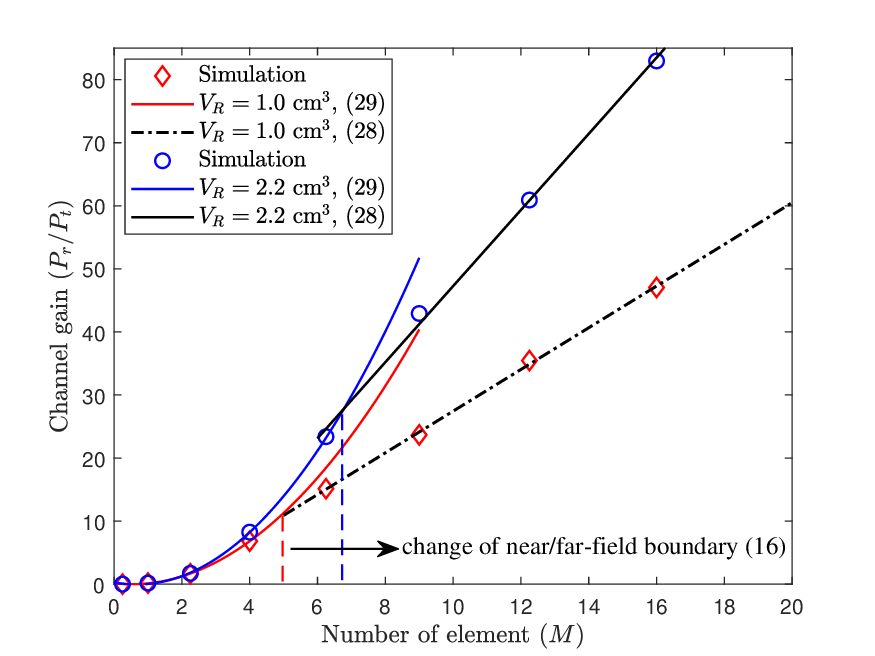}
       \caption{Power scaling laws of metasurface-based transmitting/reflecting-only RIS with receivers of different sizes.}\label{pscaling2}
       \end{center}
\end{figure*}
\subsubsection{Power Scaling Laws for Receivers with Different Sizes}
In Fig.~\ref{pscaling2}, we investigate the power scaling laws of conventional RISs for receivers with different sizes. The red markers are the simulation results for a receiver with volume $V_R = 1$ cm$^3$ while the blue markers are the results for $V_R = 1.7$ cm$^3$. As can be shown in the figure, the simulations fit well with the derived near-field and far-field power scaling laws. Within the far-field regime, both receivers exhibit linear power scaling. As derived in \eqref{scaling_far}, the blue curve has a larger slope because its larger volume. Also, the field boundary positions differ for the two receivers.

\begin{figure*}[t!]
\begin{center}
        \includegraphics[scale=0.6]{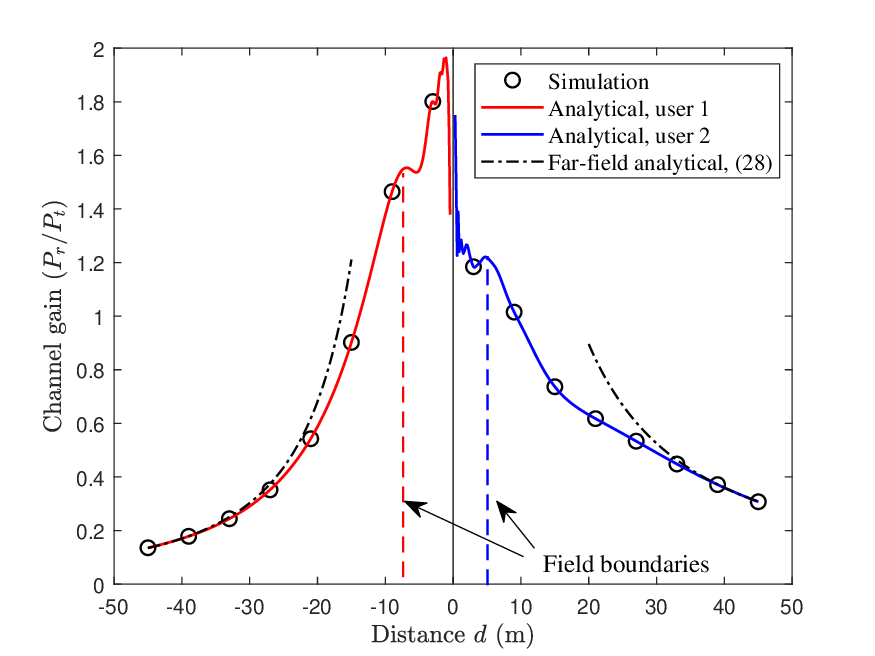}
       \caption{Simulation and analytical results for the channel gain of metasurface-based STAR-RIS.}\label{cgain1}
       \end{center}
\end{figure*}

\subsection{STAR-RIS-Aided Multi-User Scenario}

\subsubsection{Channel Gain and Field Boundary for STAR-RISs}
Fig.~\ref{cgain1} illustrates the results for the upper bound of the end-to-end channel gain. The STAR-RIS is configured to split power equally on both sides. However, the volumes of the receivers are different. User 1 (receiver with $d < 0$) has a dimension of $V_R = 0.125$ cm$^3$ and the user 2 with $d>0$ has volume $V_R = 1$ cm$^3$. The STAR-RIS is located at the center with $d=0$. As shown in the figure, within the near-fields, the channel gains of both users oscillates with the change of distance. In contrast, the far-field regimes, the channel gain follows the $d^{-2}$ dependency as derived in \eqref{scaling_far}. %Moreover, as can be observed, the field boundary for user 1 (illustrated with red dashed line) is further away. This is consistent with the analytical result in \eqref{rb} since the receiver has a larger volume, thus larger $r_b$.

\subsubsection{Comparison between the PS, REG, and SEG Strategies within Near-Field Regime}
\begin{figure*}[t!]
\begin{center}
        \includegraphics[scale=0.6]{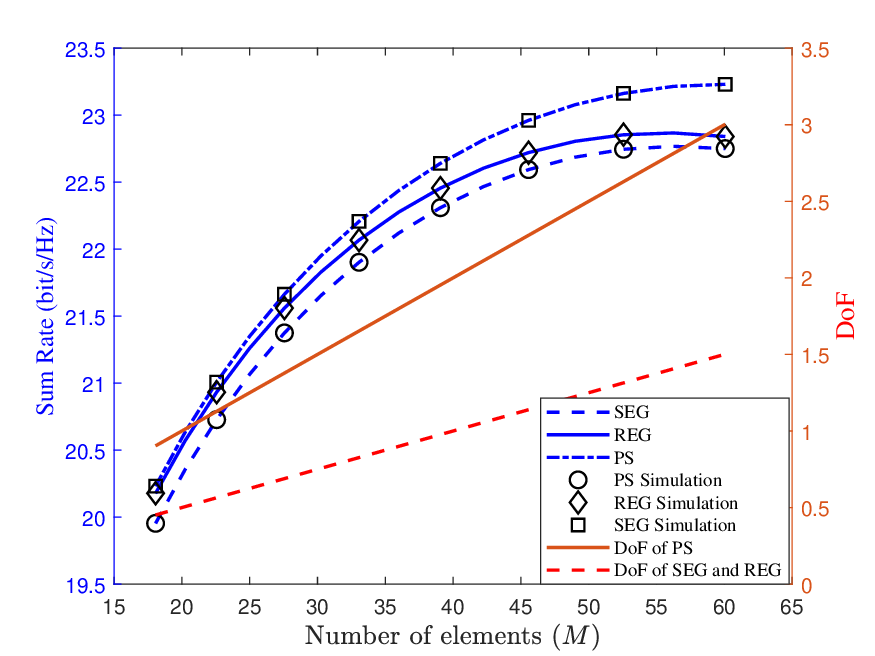}
       \caption{Simulation and analytical results for the sum rate and DoFs for two near-field users.}\label{strat}
       \end{center}
\end{figure*}

In Fig.~\ref{strat}, we investigate the sum rate performance of the three proposed configuration strategies. We consider two users, $p$ and $q$, located on different sides of the STAR-RIS. Both users are located in the x-z plane with $d = 0.5$ m to the center of the STAR-RIS (refer to Fig.~\ref{fig_two} for the coordinate system). Specifically, we assume that $x_p = -0.5$ m and $x_q = 0.3$ m. The carrier wavelength is set at $\lambda_c = 0.01$ m and the dimension of the STAR-RIS ranges from $20$ cm to $60$ cm. The dimension of the receiver is $z_R \approx 1$ cm. As can be observed in the figure, the SEG strategy outperform both REG and PS strategies, and the PS strategy has the lowest sum rate. This result is in accordance with remark \ref{eg}. Although SEG and REG strategies achieve higher sum rate, the DoFs of their corresponding channels are lower than the PS strategy. Recall that in \eqref{dof}, the DoF is proportional to the total volume of operating STAR-RIS elements. In SEG and REG, only a portion of the elements are designated to one user. As a result, for a particular user, the volume $V_T$ is smaller than the overall volume of the STAR-RIS. Thus, Fig.~\ref{strat} reveals a trade-off bettwen the sum rate and channel DoF within the near-field regime. A similar trade-off is also present in the far-field regime between the beam-forming gain and multiplexing gain of RISs or phased-array antennas~\cite{1197843}.

\begin{figure*}[t!]
\centering
\subfigure[Without STAR-RIS or window]{\label{n0}
\includegraphics[width= 2in]{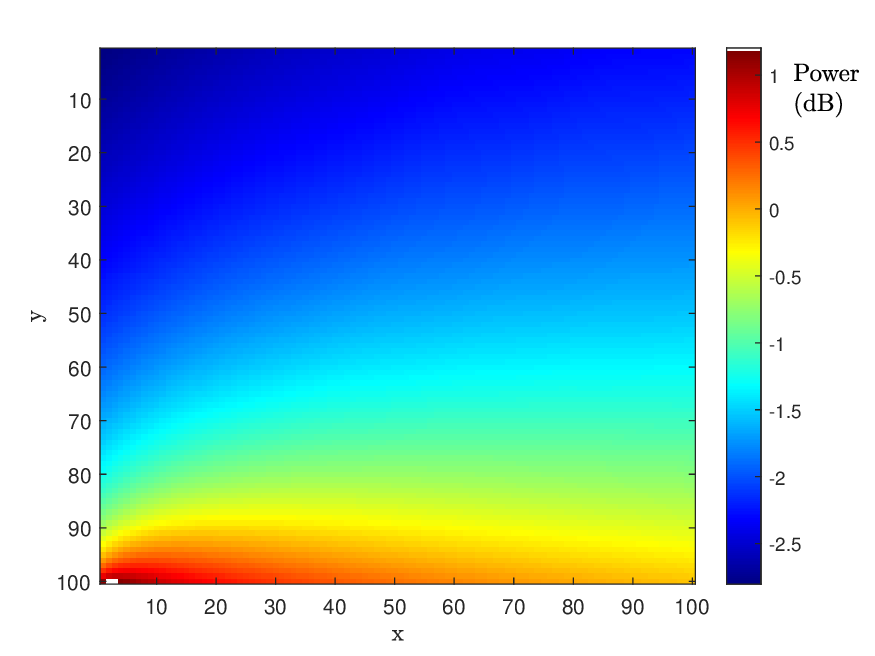}}
\subfigure[Open window]{\label{na}
\includegraphics[width= 2in]{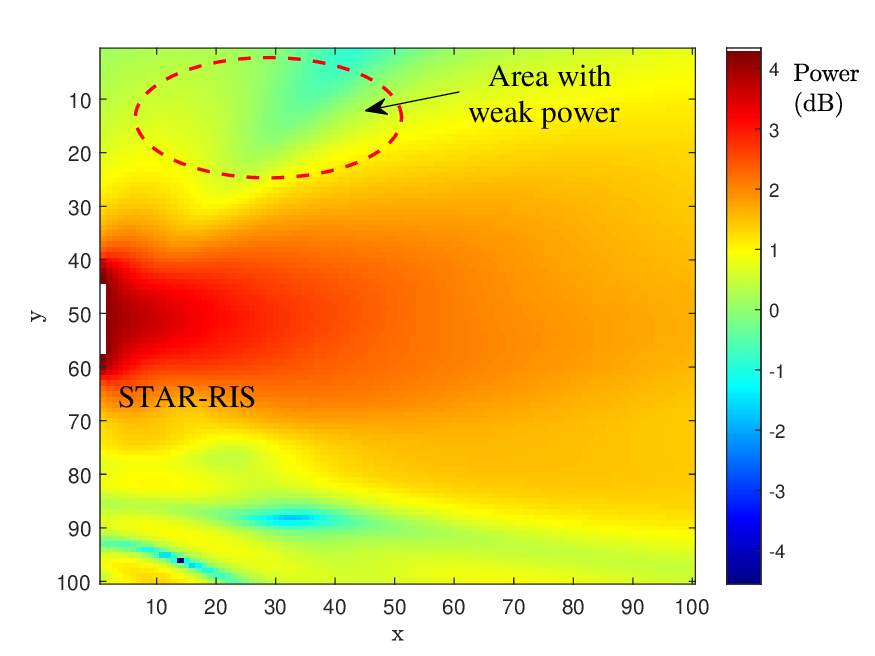}}
\subfigure[STAR-RIS]{\label{nb}
\includegraphics[width= 2in]{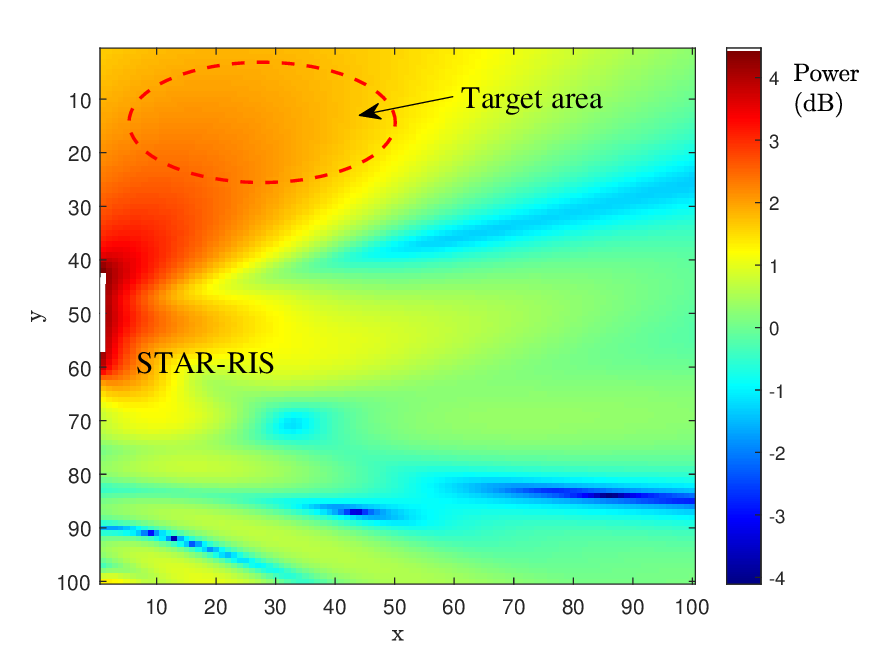}}
\caption{Simulated indoor radiation coverage.}\label{nice}
\end{figure*}
\begin{figure*}[t!]
\begin{center}
        \includegraphics[scale=0.6]{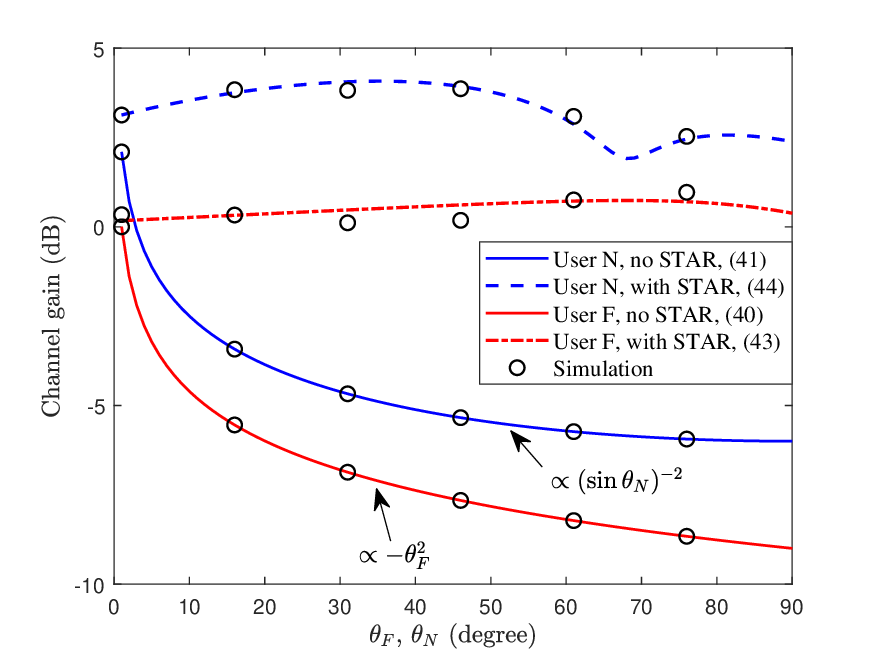}
       \caption{Simulation and analytical results for the channel gains of user $F$ and user $N$.}\label{case_deg}
       \end{center}
\end{figure*}
\subsection{STAR-RIS within Hybrid Near-Field and Far-Field Regimes}
\subsubsection{Radiation Pattern for the Indoor User}

In Fig.~\ref{nice} we plot the radiation pattern for the indoor space under the setting of the case study. Fig.~\eqref{n0} shows the power distribution without the STAR-RIS deployed. The radiation pattern agrees with the derived analytical result given in \eqref{hq}. The power falls off with $1/\sin^2\theta_N$, where $\theta_N$ is angle between the x-axis and the field-point $N$. Thus, the power is the lowest near the top-left corner of the room. Fig.~\eqref{na} shows the power distribution if a square window with size $L = 0.5$ m is opened on the center of the wall. The open window performs no passive beamforming and only lets the signal penetrates the wall within the area of $L^2 = 0.25$ m$^2$. As can be observed, the signal power is significantly better near the center of the room. However, near the top-left and bottom-left of the room, there are still regions where the signal power falls under $0$ dB. Without the passive beamforming of STAR-RIS, the signal power within these regions cannot be improved because the geometrical setting of the room is fixed.
In Fig.~\eqref{nb}, we show the power distribution in which the STAR-RIS is configured to improved the signal power near the top-left of the room. It can be observed that the STAR-RIS improves the received power within the area of interest by about $5$ dB compared to the case without STAR-RIS and $3$ dB compared to the case with an open window (without passive STAR-RIS beamforming). Furthermore, by employing the three proposed STAR-RIS configuration strategies,it is also achievable to improve the signal power at multiple locations simultaneously.

\subsubsection{Channel Gains for the Outdoor and Indoor Users}
In Fig.~\ref{case_deg}, we investigate the channel gains for user $F$ amd $N$ as a function of the angles of $\theta_F$ and $\theta_N$ (as illustrated in Fig.~\ref{system_case}). The indoor and outdoor channel gains with and without deploying the STAR-RIS are compared where the STAR-RIS adopts the PS strategy. As can be observed, the channel gains for both users without STAR-RIS decreased significantly with $\theta_F$ and $\theta_N$, as derived in \eqref{hp} and \eqref{hq}. Specifically, the channel gain for the indoor user $N$ decreases with $(\sin\theta_N)^{-2}$ and the channel gain for the outdoor user $F$ decreases with $-\theta_F^2$. The channel gains for both users are significantly improved with the aid of STAR-RIS. With the STAR-RIS deployed, the undesired angle-dependency is removed and the channel gains for both users remain high (compared to the case where $\theta_{F/N}=0$ but without STAR-RIS) within the entire angle range of $(0^\circ,90^\circ)$.

\section{Conclusions}
In this paper, a channel model based on Green's function method were proposed for investigating the performance limit of metasurface-based RISs and STAR-RISs. Instead of modeling the RIS elements with the transmission and/or reflection coefficients, we used the distribution of the induced electric currents within the metasurface-based RIS.
We also gave revealed how transmitting-only RISs, reflecting-only RISs and STAR-RISs can be achieved by configuring the distribution of the induced electric current.
For the single-user scenario with transmitting/reflecting-only RISs, the upper bound of the end-to-end channel gain was derived by choosing the current distribution that is optimized for the receiver. In addition, the position of the near-field and far-field boundary, the maximum DoF of the channel, and the power scaling law were derived. It was shown that the size of RIS, the carrier signal frequency, and the size of the receiver all affect the above performance metrics. 
For the multi-user scenario with STAR-RISs, we proposed three configuration strategies, i.e, the PS, SEG, and REG strategies. Closed-form expressed were derived for the channel gains for all three strategies. By comparing the results, it was shown that the SEG performs better than REG, especially within the near-field regime. 
We further analyzed the performance of STAR-RIS in the hybrid near-field and far-field regimes.
Specifically, we conducted study where the STAR-RIS assists the transmission of the signal from outdoor to indoor. Analytical and numerical results shown that the STAR-RIS significantly improved the signal power in the indoor ``blind zone''. The results obtained in this paper confirm the effectiveness of metasurface-based RIS-aided and STAR-RIS-aided wireless communication.

% However, there are several future research problems that require further investigation, including the channel estimation for near-field users, deployment design for STAR-RIS within indoor spaces, and the joint optimization for the active beamforming at BS and passive beamforming at STAR-RIS. 

\input{appendix}

\bibliographystyle{IEEEtran}
\bibliography{mybib}

\end{document}

%% file: appendix.tex
\begin{appendices}

\renewcommand{\theequation}{A.\arabic{equation}}
\setcounter{equation}{0}
\section{Proof of \textbf{Theorem~\ref{t_ps}}}\label{ap_c}
We start from the formulation in \eqref{hab}, and substituting the current distribution \eqref{jy2} and the approximated kernel function $K_m^{(q)}(\mathbf{r}'_1,\mathbf{r}'_2) \approx \frac{\beta^2V_R}{(4\pi r^q_m)^2}e^{-jk\Tilde{z}^{q}_1}e^{+jk\Tilde{z}^{q}_2}$ into \eqref{hab}, we have the following:
\begin{align}
    |h^{(q)}|^2 &=  \frac{DA_T}{4\pi d^2}\int_{V_T} J^*_y(\mathbf{r}'_1) \int_{V_T}
 \sum_m   \left[ \frac{\beta^2 V_R}{(4\pi r^q_m)^2}e^{-jkz^q_1}e^{-jkz^q_2} \right]
    J_y(\mathbf{r}'_2)\mathrm{d^3}\mathbf{r}'_1 \mathrm{d^3}\mathbf{r}'_2,\\ \label{b2}
    &= \frac{DA_T}{4\pi d^2}\int_{V_T}\sum_{p=1}^U e^{jkz^p_1}  \int_{V_T}
 \sum_m   \left[ \frac{\beta^2 V_R}{(4\pi r^q_m)^2}e^{-jkz^q_1}e^{jkz^q_2} \right]\sum_{p'=1}^U e^{-jkz^{p'}_2}/V_mA_m \mathrm{d}V_T \mathrm{d}V_T.
\end{align}
We first simplify the integration for $z_2$ within $V_T$. For each term in the summation of $m$, if $p'=q$ then the two exponential cancel out. And if $p' \neq q$, the we have the following integration:
\begin{equation}\label{b3}
    \int_{V_T} e^{jk(-z_2^q+z_2^{p'})} \mathrm{d}V_T
    \myeq \int_{V_T} e^{jk(-z_2^q+\cos{\alpha_m^{pq}}z_2^q)}S_m \mathrm{d}z^q_2 = \frac{2\sin\left({k(1-\cos\alpha_m^{pq})}\right)}{k(1-\cos\alpha_m^{pq}))}V_m,
\end{equation}
where in equality (a), $z_2^{p'}$ is transformed into the coordinate of centered at element $m$ with the $z$-axis pointing towards user $q$, $\alpha_m^{pq}$ is the angle between user $p$ and user $q$ from the perspective of the $m$th element, and $S_m$ is the aperture area of the $m$th element. 
The integration of $z_1$ in \eqref{b2} can be evaluated similarly. Thus, by substituting \eqref{b3} into \eqref{b2}, we have:
\begin{equation}
    |h^{(q)}|^2 =  \frac{DA_T}{4\pi d^2}\sum_m^M \frac{\beta^2 V_mV_T}{(4\pi r^q_m)^2 A_m} \left(1+\frac{2\sin\left({k(1-\cos\alpha_m^{pq})}\right)}{k(1-\cos\alpha_m^{pq}))} \right)^2.
\end{equation}

\renewcommand{\theequation}{B.\arabic{equation}}
\setcounter{equation}{0}
\section{Proof of \textbf{Theorem~\ref{t_reg}}}\label{ap_d}
Again, we start from the formulation in \eqref{hab}, and substituting the current distribution for REG in \eqref{jy3} and the approximated kernel function into \eqref{hab}, we have the following:
\begin{align}
    |h^{(p)}|^2 &=  \frac{DA_T}{4\pi d^2}\int_{V_T} J^*_{y,m}(\mathbf{r}'_1) \int_{V_T}
 \sum_m   \left[ \frac{\beta^2 V_R}{(4\pi r^p_m)^2}e^{-jkz^p_1}e^{-jkz^p_2} \right]
    J_{y,m}(\mathbf{r}'_2)\mathrm{d^3}\mathbf{r}'_1 \mathrm{d^3}\mathbf{r}'_2.
\end{align}
For the integration for $\mathbf{r}'_2$ within element $m$, the value depends on the element's grouping. 
For the elements assigned for user $p$, the focusing factor in the kernel function is cancelled by $J_{y,m}(\mathbf{r}'_2)$. However, for other elements, the integrated value yields a similar result as in \eqref{b3}. 
Thus, by evaluating the integral within each volume $(\Delta V_T)_i$ the we have:
\begin{equation}
    \int_{V_T}
 \sum_i^{N^{(p)}}   \left[ \frac{V_R}{(4\pi r^p_i)^2}e^{-jkz^p_2} \right]
    J_{y,m}(\mathbf{r}'_2) \mathrm{d^3}\mathbf{r}'_2
    =\sum_i \frac{\beta V_R\sqrt{V_m}}{(4\pi r^p_i)^2}\left(M^{(p)}+M^{(q)}\frac{2\sin\left({k(1-\cos\alpha_m^{pq})}\right)}{k(1-\cos\alpha_m^{pq}))}\right),
\end{equation}
where $M^{p}_i$ denote the number of element assigned to user $p$ within $(\Delta V_T)_i$.
Thus, the overall integration can be calculated as follows:
\begin{equation}
    |h^{(q)}|^2 =  \frac{DA_T}{4\pi d^2}\sum_i \frac{\beta V_RV_m}{(4\pi r^p_i)^2}\left(M^{(p)}+M^{(q)}\frac{2\sin\left({k(1-\cos\alpha_m^{pq})}\right)}{k(1-\cos\alpha_m^{pq}))}\right)^2,
\end{equation}
\end{appendices}